\documentclass[aps,prb,twocolumn,nofootinbib,floatfix,superscriptaddress]{revtex4-2}
\usepackage{amsmath,mleftright,mathtools}
\usepackage{bm}
\usepackage{stix,microtype}
\usepackage{graphicx}
\usepackage{xspace}
\usepackage{units}
\usepackage[colorlinks,linkcolor=blue,citecolor=blue,urlcolor=blue]{hyperref}
\usepackage[dvipsnames]{xcolor}
\usepackage{cleveref}
\usepackage{array}   
\newcolumntype{L}{>{$}c<{$}} 
\newcolumntype{C}{>{$}c<{$}} 
\usepackage{dcolumn}
\usepackage{bbm}

\makeatletter
\newcommand{\subalign}[1]{%
  \vcenter{%
    \Let@ \restore@math@cr \default@tag
    \baselineskip\fontdimen10 \scriptfont\tw@
    \advance\baselineskip\fontdimen12 \scriptfont\tw@
    \lineskip\thr@@\fontdimen8 \scriptfont\thr@@
    \lineskiplimit\lineskip
    \ialign{\hfil$\m@th\scriptstyle##$&$\m@th\scriptstyle{}##$\crcr
      #1\crcr
    }%
  }
}

\newcommand{\be}{\begin{equation}}
\newcommand{\ee}{\end{equation}}
\newcommand{\bea}{\begin{eqnarray}}
\newcommand{\eea}{\end{eqnarray}}

\def\a{\alpha}
\def\b{\beta}
\def\g{\gamma}
\def\G{\Gamma}
\def\d{\delta}
\def\D{\Delta}
\def\e{\epsilon}

\def\th{\theta}
\def\Th{\Theta}

\def\m{\mu}
\def\n{\nu}

\def\p{\pi}
\def\P{\Pi}
\def\r{\rho}
\def\s{\sigma}
\def\S{\Sigma}

\def\f{\phi}

\def\w{\omega}
\def\W{\Omega}
\def\q{\psi}
\def\Q{\Psi}

\def\bgf{\mbox{\boldmath $\phi$}}

\def\ble{{\mathbf e}}

\def\blk{{\mathbf k}}

\def\bln{{\mathbf n}}

\def\blp{{\mathbf p}}
\def\blq{{\mathbf q}}
\def\blr{{\mathbf r}}

\def\blx{{\mathbf x}}

\def\blA{{\mathbf A}}

\def\blP{{\mathbf P}}

\def\blU{{\mathbf U}}

\def\callA{\mbox{$\mathcal{A}$}}

\def\callW{\mbox{$\mathcal{W}$}}

\def\de{\partial}
\def\iif{\infty}
\def\bra{\langle}
\def\ket{\rangle}

\def\Re{{\rm Re}}

\def\iu{{\rm i}}
\def\1op{\hat{\mathbbm{1}}}
\def\nn{\nonumber}

\keywords{}
\begin{document}
\title{Semiconductor Electron-Phonon Equations: a Rung Above 
Boltzmann in the Many-Body Ladder}
\author{Gianluca Stefanucci}
\affiliation{Dipartimento di Fisica, Universit{\`a} di Roma Tor Vergata, Via della Ricerca Scientifica 1,
00133 Rome, Italy}
\affiliation{INFN, Sezione di Roma Tor Vergata, Via della Ricerca Scientifica 1, 00133 Rome, Italy}
\author{Enrico Perfetto}
\affiliation{Dipartimento di Fisica, Universit{\`a} di Roma Tor Vergata, Via della Ricerca Scientifica 1,
00133 Rome, Italy}
\affiliation{INFN, Sezione di Roma Tor Vergata, Via della Ricerca Scientifica 1, 00133 Rome, Italy}
\date{\today}

\begin{abstract}  	
Starting from the {\em ab initio}  many-body theory of electrons and 
phonons, we go through a series of well defined 
simplifications to derive a set of coupled equations of motion for the 
electronic occupations and polarizations, nuclear displacements as 
well as phononic occupations and coherences. These are the 
semiconductor electron-phonon equations (SEPE), sharing the same 
scaling with system size and propagation time as the Boltzmann 
equations. At the core of the SEPE is the {\em mirrored} Generalized 
Kadanoff-Baym ansatz (GKBA) for the Green's 
functions, an alternative to the standard GKBA which we show to lead to unstable 
equilibrium states. 
The SEPE treat coherent and incoherent degrees of 
freedom on equal footing, widen the scope of the semiconductor Bloch 
equations and Boltzmann 
equations, and reduce to 
them under additional simplifications. The new features of the SEPE
pave the way for first-principles studies of phonon squeezed states 
and coherence effects in time-resolved absorption and diffraction 
experiments.

\end{abstract}
\maketitle

\section{Introduction}

In a recent work we have laid down the {\em ab initio} many-body theory of 
electrons and phonons~\cite{stefanucci_in-and-out_2023}, and derived 
the Kadanoff-Baym equations (KBE)~\cite{kadanoff1962quantum,svl-book} 
for the electronic and phononic Green's 
functions. Solution of the KBE would provide us with a detailed understanding
of the system's dynamics as we could extract electronic occupations 
and polarizations, nuclear displacements, phononic occupations and 
coherences as well as electronic and phononic spectral functions, 
life-times, quasi-particle renormalizations,
satellites, etc. Unfortunately, the unfavorable scaling 
with the propagation 
time~\cite{dahlen_solving_2007,myohanen_a-many-body_2008,myohanen_kadanoff-baym_2009,pva.2009,pva.2010,schuler_time-dependent_2016,schuler_nessi_2020}, which is at least cubic,  continues to render the 
KBE challenging for {\em ab initio} simulations, 
though some progress has been made~\cite{kaye_low-rank_2021,meirinhos_adaptive_2022,dong_excitations_2022}.

For crystals with a finite quasi-particle 
gap in the one-particle spectrum, like 
semiconductors or insulators, alternative (and approximate) 
theoretical frameworks to the KBE include the 
semiconductor Bloch equations (SBE)~\cite{haug_quantum_1994} and the 
Boltzmann equations (BE).
At clamped nuclei both these frameworks can be derived from the KBE 
through well identifiable simplifications; in other words, we 
precisely understand what is missing.
Orthodox derivations of the SBE 
are based on the cluster expansion~\cite{kira_many-body_2006}, while 
the Wigner 
function~\cite{kadanoff1962quantum,danielewicz_quantum_1984} 
or the Fermi golden 
rule~\cite{ziman_electrons_1960,ponce_toward_2018,sadasivam_theory_2017} are 
usually invoked to derive 
the BE.
Alternatively one can use low-order diagrammatic expansions and 
the so called 
Generalized Kadanoff-Baym Ansatz (GKBA) for the electronic Green's 
function~\cite{lipavsky_generalized_1986}, which allows for mapping 
the KBE onto a single equation for the one-electron density matrix, 
see below. 
The situation is not as 
clear when electrons and phonons are treated on equal footing. 
The reason is twofold: 
Firstly the KBE for electrons and phonons have been 
only recently established~\cite{stefanucci_in-and-out_2023}. 
Secondly, a GKBA for phonons has 
been proposed only a couple of years ago~\cite{karlsson_fast_2021,pavlyukh_time-linear_2022}.
At present, both the SBE and BE 
for electrons {\em and} phonons
must be regarded as semi-empirical frameworks since 
they are not derived from the {\em ab initio} Hamiltonian.

We here climb down the many-body ladder starting from the 
highest rung, i.e., the {\em ab initio} KBE. 
We present an alternative ansatz to the GKBA, which we name 
the {\em mirrored} GKBA (MGKBA). The motivation for introducing the 
MGKBA arises from the observation that combining the GKBA with the 
Markov approximation results in unphysical outcomes and an unstable 
equilibrium state. Through a series of well-defined 
simplifications we 
then derive a set of coupled equations for the electronic occupations and 
polarizations, nuclear displacements, and phononic occupations and 
coherences, which we name the {\em semiconductor electron-phonon 
equations} (SEPE).  The SEPE scale 
like the SBE and BE with propagation time and system size. 
Their unique feature over the 
SBE and BE is a consistent treatment of 
phononic occupations and coherences as well as the
inclusion of the renormalization of the electronic quasi-particle energies 
induced by the nuclear displacements. The former aspect opens the 
door to studies of phonon squeezed states in optically excited 
semiconductors~\cite{garrett_vacuum_1997,johnson_directly_2009,benatti_generation_2017,lakehal_detection_2020}. 
The latter aspect is relevant 
for capturing the coherent modulation of time-resolved optical 
spectra of resonantly pumped semiconductors~\cite{trovatello_strongly_2020,li_single-layer_2021,mor_photoinduced_2021,jeong_coherent_2016,sayers_strong_2023}.
We finally elucidate the 
theoretical underpinnings of the SBE and BE, demonstrating how they emerge from
the SEPE when additional simplifications are made.

The paper is organized as follows. In Sections~\ref{aihamsec} and~\ref{aikbesec} we 
briefly revisit the {\em ab initio} many-body theory of electrons and 
phonons~\cite{stefanucci_in-and-out_2023}. In Section~\ref{ephdmsec} 
we introduce the electronic and phononic density matrices, and derive 
their exact equations of motion in terms of electronic and phononic 
self-energies. The GKBA and MGKBA is discussed in 
Section~\ref{sgkbasec} while self-energies and screening effects 
are presented in Section~\ref{sesec}. Section~\ref{sepesec} is the 
core of this work; it contains the derivation of the SEPE and a 
characterization of the solutions in the long-time limit. 
How to recover the SBE and BE is the topic of Section~\ref{recoverysec}.
A summary of the main findings and an outlook on future 
applications are drawn in Section~\ref{conclsec}.

\section{Ab initio Hamiltonian for electrons and phonons}
\label{aihamsec}

Let us consider a semiconductor or an insulator 
and assign a suitable 
basis to expand the electronic field operators and the nuclear 
displacement operators:
\begin{align}
\hat{\q}(\blx)&=\sum_{\blk\m}\Q_{\blk\m}(\blx)\hat{d}_{\blk\m},
\label{fieldop}
\\
\hat{U}_{\bln s,j}&=\frac{1}{\sqrt{M_{s}N}}
\sum_{\blq}e^{i\blq\cdot\bln}\,
\sum_{\n}
e^{\a}_{s,j}(\blq)
\;\hat{U}_{\blq \a}.
\label{displdecnm}
\end{align}
In Eq.~(\ref{fieldop})
the Bloch wavefunctions $\Q_{\blk\m}(\blx=\blr\s)$ (position $\blr$ and 
spin $\s$) are the eigenfunctions of a one-particle Hamiltonian
with the periodicity of 
the crystal, e.g., the Kohn-Sham Hamiltonian (hence the index $\m$ can 
be thought of as a band index). In Eq.~(\ref{displdecnm}) 
$\hat{U}_{\bln s,j}$ is the displacement along direction $j=x,y,z$ 
of nucleus $s$ (with mass $M_{s}$) in cell $\bln$, the total number 
of cells being $N$. 
The unit vectors $\ble^{\a}(\blq)$, with components 
$e^{\a}_{s,j}(\blq)$, 
form an orthonormal basis for each $\blq$. 
Although not necessary in this section, we 
can already choose these vectors to be the {\em normal modes} of the 
Born-Oppenheimer (BO) energy. The hermiticity of the operators 
$\hat{U}_{\bln s,j}$ implies that $\hat{U}_{\blq \a}=\hat{U}_{-\blq 
\a}^{\dag}$ and $\ble^{\a\ast}(-\blq)=\ble^{\a}(\blq)$. 

Let $\hat{P}_{\blq \a}=\hat{P}_{-\blq \a}^{\dag}$ be the conjugate 
momentum of $\hat{U}_{\blq \a}$, i.e., $[\hat{U}_{\blq 
\a},\hat{P}_{\blq' \a'}^{\dag}]=\d_{\blq,\blq'}\d_{\a\a'}$. 
The {\em ab initio} Hamiltonian for electrons and nuclei in the harmonic 
approximation can be written as (atomic units are used throughout 
this work)~\cite{stefanucci_in-and-out_2023}
\begin{align}
\hat{H}=\hat{H}_{0,e}+\hat{H}_{0,ph}+
\hat{H}_{e-e}+\hat{H}_{e-ph},
\label{el-phonham3}
\end{align}
where
\begin{subequations}
\begin{align}
\hat{H}_{0,e}&=\sum_{\blk\m\m'}h_{\m\m'}(\blk)\hat{d}^{\dag}_{\blk\m}\hat{d}_{\blk\m'},
\label{h0es}
\\
\hat{H}_{0,ph}&=\frac{1}{2}\sum_{\blq\a\a'}
\big(\hat{\blU}_{\blq\a}^{\dag},\hat{\blP}_{\blq\a}^{\dag}\big)
\left(\!\!\begin{array}{cc}
	K_{\a\a'}(\blq) & 0 \\ 0 & \d_{\a\a'} 
\end{array}\!\!\right)
\left(\begin{array}{cc}
	\hat{\blU}_{\blq\a'} \\ \hat{\blP}_{\blq\a'} 
\end{array}\right)
\nn\\
&-\sum_{\blk\m\m'}\sum_{\a}
\r^{\rm eq}_{\blk\m'\m}
g_{{\mathbf 0}\a,\m\m'}(\blk)\,
\hat{\blU}_{{\mathbf 0}\a},
\label{h0phs}
\\
\hat{H}_{e-e}&=
\frac{1}{2}\sum_{\substack{\blk\blk'\blq\\ \m\m'\n\n'}}
\!\!v_{\blk+\blq\m\,\blk'-\blq\n'\,\blk'\n\,\blk\m'}
\hat{d}^{\dag}_{\blk+\blq\m}\hat{d}^{\dag}_{\blk'-\blq\n'}
\hat{d}_{\blk'\n}\hat{d}_{\blk\m'},
\label{hees}
\\
\hat{H}_{e-ph}&=
\sum_{\blk\m\m'}\sum_{\blq\a}
\hat{d}^{\dag}_{\blk\m}\hat{d}_{\blk-\blq\m'}
g_{-\blq\a,\m\m'}(\blk)\,
\hat{\blU}_{\blq\a}.
\label{hephs}
\end{align}
\label{hcomp}
\end{subequations}
In Eq.~(\ref{h0es}), 
$h_{\m\m'}(\blk)=\bra\blk\m|\frac{\hat{\blp}^{2}}{2}+V(\hat{\blr})|\blk\m'\ket$ 
is the matrix element of the one-electron Hamiltonian, $V(\blr)$ 
being the potential generated by the nuclei in their equilibrium 
positions. Equation~(\ref{h0phs}) is the Hamiltonian of the bare 
phonons, with $K_{\a\a'}(\blq)$ the elastic tensor, 
$g_{-\blq\a,\m\m'}(\blk)$ the electron-phonon ($e$-$ph$) coupling and 
$\r^{\rm eq}_{\blk\m'\m}=\bra 
\hat{d}^{\dag}_{\blk\m}\hat{d}_{\blk\m'}\ket$ the equilibrium 
one-electron density matrix. 
As pointed out in Ref.~\cite{stefanucci_in-and-out_2023} the second 
line of Eq.~(\ref{h0phs}) plays a pivotal role in proving  that the time-derivative of the 
nuclear momenta vanish in equilibrium.
The explicit form of the 
elastic tensor is not important here, rather it is relevant to 
say that adding 
to it the equilibrium phononic self-energy $\P_{\blq\a\a'}(\w)$ in the 
clamped-nuclei approximation evaluated at $\w=0$ we have the exact identity 
(in the basis of the BO normal 
modes)~\cite{feliciano_electron-phonon_2017,baroni_phonons_2001}
\begin{align}
K_{\a\a'}(\blq)+\P_{\blq\a\a'}(\w=0)=\d_{\a\a'}\w^{2}_{\blq\a},
\label{exactid}
\end{align}
where $\w^{2}_{\blq\a}$ are the eigenvalues of the Hessian of the BO 
energy. The $e$-$ph$ coupling is defined as
\begin{align}
g_{-\blq\a,\m\m'}(\blk)=\bra\blk\m|\left.\frac{\de V(\hat{\blr})}
{\de U_{\blq\a}}\right|_{\blU=0}|\blk+\blq\m'\ket=
g^{\ast}_{\blq\a,\m'\m}(\blk-\blq).
\label{ephcoupprop}
\end{align}
The electron-electron ($e$-$e$) interaction is described by 
Eq.~(\ref{hees}), with 
$v_{\blk+\blq\m\,\blk'-\blq\n'\,\blk'\n\,\blk\m'}=\bra
\blk+\blq\m\,\blk'-\blq\n'|v(\hat{\blr},\hat{\blr}')|\blk'\n\,\blk\m'\ket$
the Coulomb scattering amplitudes. 
The $e$-$ph$ interaction is accounted for by Eq.~(\ref{hephs}). 

We 
are interested in studying the dynamics of the system photoexcited by 
an external driving field
\begin{align}
\hat{H}_{\rm drive}(t)=\sum_{\blk\m\m'}\W_{\blk\m\m'}(t)\hat{d}_{\blk\m}^{\dag}
\hat{d}_{\blk\m'},
\label{hdrive}
\end{align}
with Rabi frequencies
\begin{align}
\W_{\blk\m\m'}(t)\equiv \frac{1}{2c}\bra\blk\m|\hat{\blp}\cdot 
\blA(\hat{\blr},t)+\blA(\hat{\blr},t)\cdot\hat{\blp}
+\frac{1}{c}\blA^{2}(\hat{\blr},t)|\blk\m'\ket.
\end{align}
We are here implicitly assuming that the  
driving field does not break the lattice periodicity. 

\section{Ab initio Kadanoff-Baym equations for electrons and phonons}
\label{aikbesec}

We find it convenient to arrange the 
displacement and momentum operators into a two-dimensional 
vector
\begin{align}
\hat{\bgf}_{\blq\a}=\left(\begin{array}{c} 
\hat{\f}_{\blq\a}^{1}\\\hat{\f}_{\blq\a}^{2}
\end{array}\right)=
\left(\begin{array}{c}
\hat{U}_{\blq\a}\\\hat{P}_{\blq\a}
\end{array}\right).
\label{phicomp}
\end{align}
The {\em ab initio} KBE are coupled 
integro-differential equations 
for the electronic greater and lesser Green's functions (GFs) 
\begin{subequations}
\begin{align}
G^{>}_{\blk\m\m'}(t,t')&=-i\bra\hat{d}_{\blk\m,H}(t)	
\hat{d}_{\blk\m',H}^{\dag}(t')\ket,
\\
G^{<}_{\blk\m\m'}(t,t')&=i\bra\hat{d}_{\blk\m',H}^{\dag}(t')
\hat{d}_{\blk\m,H}(t)\ket,
\end{align}
\end{subequations}
and phononic greater and lesser GF ($i,j=1,2$):
\begin{subequations}
\begin{align}
D^{ij,>}_{\blq\a\a'}(t,t')&=-i\bra\D\hat{\f}^{i}_{\blq\a,H}(t)	
\D\hat{\f}^{j}_{-\blq\a',H}(t')\ket,
\\
D^{ij,<}_{\blq\a\a'}(t,t')&=-i\bra\D\hat{\f}^{j}_{-\blq\a',H}(t')
\D\hat{\f}^{i}_{\blq\a,H}(t)\ket,
\end{align}
\end{subequations}
where 
$\D\hat{\f}^{j}_{\blq\a}=\hat{\f}^{j}_{\blq\a}-\bra\hat{\f}^{j}_{\blq\a}\ket$ 
is the fluctuation operator.
In these definitions the operators are in the Heisenberg picture with 
respect to the time-dependent Hamiltonian $\hat{H}+\hat{H}_{\rm 
drive}(t)$. The KBE read (in matrix form)~\cite{stefanucci_in-and-out_2023}
\begin{subequations}
\begin{align}
\Big[i\frac{d}{dt}-h(\blk,t)
&-\sum_{\a}g_{{\mathbf 0}\a}(\blk)U_{{\mathbf 0}\a}(t)\Big]
G_{\blk}^{\lessgtr}(t,t')
\nn \\
&=\left[
\S_{\blk}^{\rm R}\cdot G_{\blk}^{\lessgtr}+
\S_{\blk}^{\lessgtr}\cdot G_{\blk}^{\rm A}
\right](t,t'),
\label{eomG<>1}
\end{align}
\begin{align}
\Big[i\callA
\frac{d}{dt} -&Q(\blq)
\Big]
D^{\lessgtr}_{\blq}(t,t')=\left[
\P^{\rm R}_{\blq}\cdot D^{\lessgtr}_{\blq}+
\P^{\lessgtr}_{\blq}\cdot D^{\rm A}_{\blq}
\right](t,t'),
\label{eomD<>1}
\end{align}
\label{eomXlessgtr}
\end{subequations}
where we use the symbol ``~$\cdot$~''   to denote time-convolutions.
In Eqs.~(\ref{eomXlessgtr}) 
\begin{align}
h_{\m\m'}(\blk,t)=h_{\m\m'}(\blk)+\W_{\blk\m\m'}(t),
\end{align}
and 
\begin{align}
\callA_{\a\a'}=\d_{\a\a'}\left(\!
\begin{array}{cc}
	0 & i \\ -i & 0 
\end{array}
\!\right)\quad,\quad
Q_{\a\a'}(\blq)&=\left(\!\begin{array}{cc}
	K_{\a\a'}(\blq) & 0 \\ 0 & \d_{\a\a'} 
\end{array}\!\right).
\end{align}
The r.h.s. of the KBE contains the 
electronic self-energy
$\S_{\blk}$ and   the 
phononic self-energy $\P_{\blq}$. As the electrons couple only to the 
nuclear displacements we have $\P_{\blq}^{ij}=\d_{i1}\d_{j1}\P_{\blq}$.
Henceforth we use the same symbol $\P_{\blq}$ to represent the 
$2\times 2$ phononic self-energy and its $(1,1)$ element; whether 
$\P_{\blq}$ is a matrix or a scalar is evident from the context.
The retarded (R) and advanced (A) correlators are defined in terms of 
the lesser and greater correlators according to 
$X^{\rm R/A}(t,t')=\d(t-t')X^{\d}(t)\pm\th(\pm t\mp 
t')[X^{>}(t,t')-X^{<}(t,t')]$, where $X^{\d}$ is the weight of a 
possible singular part of $X$.
For $G$ and $D$ the singular part is zero in all 
approximations.  
As we see later this is not the case for $\S$ and $\P$. The KBE are
coupled to the equation of motion of the nuclear displacement 
[see l.h.s. of Eq.~(\ref{eomG<>1})]~\cite{stefanucci_in-and-out_2023} 
\begin{align}
\frac{d^{2}}{dt^{2}}U_{{\mathbf 0}\a}(t)
=-\sum_{\blk\m\n}g_{{\mathbf 
0}\a,\n\m}(\blk)\D \r^{<}_{\blk\m\n}(t)
-\sum_{\a'}K_{\a\a'}({\mathbf 0})U_{{\mathbf 0}\a'}(t),
\label{sducp}
\end{align}
where $\D \r^{<}_{\blk\m\n}(t)=-iG^{<}_{\blk\m\n}(t,t)-
\r^{\rm eq}_{\blk\m\n}$.

Solving the KBE with exact self-energies yield the exact two-times 
GFs. These provide information on the dynamics of carriers, phonon 
occupations and coherences as well as spectral properties relevant to 
time-resolved ARPES and Raman experiments.
However, the time non-locality of the self-energies represents a major 
numerical obstacle for a full two-times propagation. 
The time-convolutions in the r.h.s. of Eqs.~(\ref{eomXlessgtr}) make 
any time-stepping algorithm scale at least
cubically with the propagation time.
In the following we introduce a series of simplifications 
leading to the SEPE, i.e., a couple system of ordinary differential equations for the 
electronic and phononic density matrices. 
The numerical solution of the SEPE scales linearly in time.
The semiconductor Bloch equations and the Boltzmann equations follow from 
the SEPE by making additional simplifications.

\section{Electronic and phononic density matrices}
\label{ephdmsec}

The electronic and phononic density matrices are proportional to the 
equal-time electronic and phononic GFs. We define them according to
\begin{subequations}
\begin{align}
\r^{\gtrless}_{\blk\m\m'}(t)&\equiv -i G^{\gtrless}_{\blk\m\m'}(t,t),
\label{eldenmat><def}
\\
\g^{\gtrless}_{\blq\a\a'}(t)&\equiv i \,D^{\gtrless}_{\blq\a\a'}(t,t).
\label{gammadef}
\end{align}
\end{subequations}
Using the commutation rules for the electronic and nuclear operators 
we easily find
\begin{subequations}
\begin{align}
	\r^{>}_{\blk\m\m'}(t)&=\r_{\blk\m\m'}^{<}(t)-\d_{\m\m'},
\label{commreln<>}
\\
\g^{>}_{\blq\a\a'}(t)&=\g^{<}_{\blq\a\a'}(t)+\callA_{\a\a'}.
\label{g>=g<+a}
\end{align}
\end{subequations}

The electronic density matrix 
$\r^{<}_{\blk\m\m'}(t)=\bra\hat{d}^{\dag}_{\blk\m',H}(t)\hat{d}_{\blk\m,H}(t)\ket$ 
is self-adjoint in the space of the band indices, i.e., 
$[\r^{<}_{\blk}]^{\dag}=\r^{<}_{\blk}$. The diagonal entries are 
non-negative and determine
the electronic occupations
\begin{align}
f^{\rm el}_{\blk\m}(t)\equiv \r^{<}_{\blk\m\m}(t),
\end{align}
while the off-diagonal entries provide information 
on the electronic polarization
\begin{align}
p_{\blk\m\m'}(t)\equiv \r^{<}_{\blk\m\m'}(t),
\quad \m\neq\m'.
\end{align}
Similarly 
$[\r^{>}_{\blk}]^{\dag}=\r^{>}_{\blk}$ has non-positive 
diagonal entries determining the negative of the hole occupations.

The phononic density matrix 
$\g^{ij,<}_{\blq\a\a'}$ 
is self-adjoint in the direct-product space of the normal mode indices and 
components, i.e., $[\g^{<}_{\blq}]^{\dag}=\g^{<}_{\blq}$, and 
similarly $[\g^{>}_{\blq}]^{\dag}=\g^{>}_{\blq}$.
To gain some more physical intuition on the phononic density matrix 
we write the displacements and momenta in terms of 
dressed phononic operators
\begin{subequations}
\begin{align}
\hat{U}_{\blq \a}&=\frac{1}{\sqrt{2\w_{\blq\a}}}
(\hat{b}_{\blq\a}+\hat{b}^{\dag}_{-\blq\a}),
\\
\hat{P}_{\blq \a}&=-i\sqrt{\frac{\w_{\blq\a}}{2}}
(\hat{b}_{\blq\a}-\hat{b}^{\dag}_{-\blq\a}).
\end{align}
\label{intrphonopdressed}
\end{subequations}
The commutation relations between $\hat{U}_{\blq \a}$ and 
$\hat{P}_{\blq' \a'}$ are satisfied {\em for any} $\w_{\blq\a}>0$ 
provided that 
$[\hat{b}_{\blq\a},\hat{b}^{\dag}_{\blq'\a'}]=\d_{\blq,\blq'}\d_{\a\a'}$.
For later purposes we here take the $\w_{\blq\a}$'s to be the BO 
frequencies, see Eq.~(\ref{exactid}).
The average values $U_{\blq\a}(t)=\bra\hat{U}_{\blq\a,H}(t)\ket$ and 
$P_{\blq\a}(t)=\bra\hat{P}_{\blq\a,H}(t)\ket$ is zero for all 
$\blq\neq {\mathbf 0}$ due to the fact that the lattice periodicity 
is preserved. Then, for all $\blq\neq {\mathbf 0}$ and 
for $\a=\a'$ we have
\begin{subequations}
\begin{align}
\g^{11,<}_{\blq\a\a}(t)&=\bra\D\hat{U}_{-\blq\a,H}(t)\D\hat{U}_{\blq\a,H}(t)\ket
\nn\\&=\frac{1}{2\w_{\blq\a}}\Big(f^{\rm ph}_{\blq\a}(t)+f^{\rm 
ph}_{-\blq\a}(t)+1+\Th_{\blq\a}(t)+\Th^{\ast}_{\blq\a}(t)\Big),
\end{align}
\begin{align}
\g^{22,<}_{\blq\a\a}(t)&=\bra\D\hat{P}_{-\blq\a,H}(t)\D\hat{P}_{\blq\a,H}(t)\ket
\nn\\&=\frac{\w_{\blq\a}}{2}\Big(f^{\rm ph}_{\blq\a}(t)+f^{\rm 
ph}_{-\blq\a}(t)+1-\Th_{\blq\a}(t)-\Th^{\ast}_{\blq\a}(t)]\Big),
\end{align}
\begin{align}
\g^{12,<}_{\blq\a\a}(t)&
=\bra\D\hat{P}_{-\blq\a,H}(t)\D\hat{U}_{\blq\a,H}(t)\ket
\nn\\&=\frac{i}{2}\Big(f^{\rm ph}_{\blq\a}(t)-f^{\rm 
ph}_{-\blq\a}(t)-1-\Th_{\blq\a}(t)+\Th^{\ast}_{\blq\a}(t)]\Big),
\end{align}
\label{gammamatele}
\end{subequations}
where  we introduce the phononic 
occupations 
\begin{align}
f^{\rm ph}_{\blq\a}(t)\equiv \bra
b^{\dag}_{\blq\a,H}(t)b_{\blq\a,H}(t)\ket,
\label{phonoccdef}
\end{align}
and the {\em phononic coherences}
\begin{align}
\Th_{\blq\a}(t)=\Th_{-\blq\a}(t)\equiv \bra
b_{\blq\a,H}(t)b_{-\blq\a,H}(t)\ket.
\label{phoncohdef}
\end{align}
An important property satisfied by the diagonal entries 
is
\begin{align}
\g^{ij,<}_{\blq\a\a}(t)=\g^{ji,>}_{-\blq\a\a}(t).
\label{diagpropgamma}
\end{align}
We use Eq.~(\ref{diagpropgamma}) in our subsequent derivations.

The exact equation of motion for $\r^{<}_{\blk}(t)$ can be derived by 
subtracting the lesser form of Eq.~(\ref{eomG<>1}) to its adjoint  
and then setting $t'=t$:
\begin{align}
\frac{d}{dt}&\r^{<}_{\blk}(t)+i
\big[h_{\rm qp}(\blk,t),\r^{<}_{\blk}(t)\big]
\nn\\
&=-\int^{t} \!\!dt'
\left[\S^{>}_{\blk}(t,t')G^{<}_{\blk}(t',t)-
\S^{<}_{\blk}(t,t')G^{>}_{\blk}(t',t)\right]
+{\rm h.c.},
\label{eomrhot}
\end{align}
where 
\begin{align}
h_{\rm qp}(\blk,t)= h(\blk,t)+\sum_{\a}g_{{\mathbf 
0}\a}(\blk)U_{{\mathbf 0}\a}(t)+\S^{\d}_{\blk}(t),
\end{align}
is the so called quasi-particle Hamiltonian, 
and ``h.c.'' stands for the hermitian conjugate. 
Often $\S^{\d}$ is evaluated in the Hartree plus statically screened 
exchange (HSEX)
approximation. 
Similarly, the exact equation of motion for $\g^{<}_{\blq}(t)$ can be derived by 
subtracting the lesser form of Eq.~(\ref{eomD<>1}) to its adjoint
and then setting $t=t'$:
\begin{align}
&\frac{d}{dt}\g^{<}_{\blq}(t)+i
\Big(\callA Q_{\rm qp}(\blq,t)\g^{<}_{\blq}(t)-
\g^{<}_{\blq}(t)Q_{\rm qp}(\blq,t)\callA\Big)
\nn\\
&=\callA\int^{t} \!\!dt'\left[\P^{>}_{\blq}(t,t') D^{<}_{\blq}(t',t)-
\P^{<}_{\blq}(t,t') D^{>}_{\blq}(t',t)\right]
+{\rm h.c.},
\label{eomd<tt}
\end{align}
where
\begin{align}
Q_{\rm qp}(\blq,t)\equiv Q(\blq)+\P^{\d}_{\blq}(t),
\label{qpQ}
\end{align}
is the quasi-phonon, or dressed-phonon, Hamiltonian. A physically 
sensible approximation to $\P^{\d}$ is discussed in 
Section~\ref{bosgkbasec}.

\section{Generalized Kadanoff-Baym Ansatz and its mirrored form}
\label{sgkbasec}

To close the equations of motion of the density matrices 
we have to transform the time off-diagonal 
$G^{\gtrless}$ and $D^{\gtrless}$ into functionals
of $\r^{<}$ and $\g^{<}$. 
In this way the r.h.s. in 
Eqs.~(\ref{eomrhot}) and (\ref{eomd<tt}) 
became functionals of $\r^{<}$ and $\g^{<}$ since 
$\S^{\gtrless}$ and $\P^{\gtrless}$ are functionals of 
$G^{\gtrless}$ and  $D^{\gtrless}$.

\subsection{Electrons}

In the mid-1980's Lipavsk\'y et al.~\cite{lipavsky_generalized_1986} 
proposed an ansatz for the $G^{\gtrless}$-functional. In 
essence the idea is to manipulate and then modify the following 
exact relation for the noninteracting GF $G_{0}$~\cite{svl-book}:
\begin{align}
G_{0,\blk}^{<}(t,t')=
G_{0,\blk}^{\rm R}(t,0)
G_{0,\blk}^{<}(0,0)
G_{0,\blk}^{\rm A}(0,t'),
\end{align}
where 
\begin{align}
G_{0,\blk}^{\rm R}(t,t')=[G_{0,\blk}^{\rm A}(t',t)]^{\dag}=
-i\th(t-t')T \big\{e^{-i\int_{t'}^{t}d\bar{t}\,h(\blk,\bar{t})}\big\},
\label{G0ret}
\end{align}
$T$ being the 
time-ordering operator. Using the 
group property 
$G_{0,\blk}^{\rm R}(t,0)=i 
G_{0,\blk}^{\rm R}(t,t')G_{0,\blk}^{\rm 
R}(t',0)$ for all $t>t'>0$, and the like for the 
advanced GF, we find
\begin{align}
G_{0,\blk}^{<}(t,t')=
-G_{0,\blk}^{\rm R}(t,t')\r_{0,\blk}^{<}(t')+
\r_{0,\blk}^{<}(t)G_{0,\blk}^{\rm 
A}(t,t'),
\label{gkbastep1}
\end{align}
where $\r_{0,\blk}^{<}(t)\equiv -i G_{0,\blk}^{<}(t,t)$. 
An identical relation holds for $G_{0,\blk}^{>}(t,t')$ 
provided that we replace $\r_{0,\blk}^{<}$ with 
$\r_{0,\blk}^{>}$.
The {\em Generalized Kadanoff-Baym 
Ansatz}\index{GKBA}\index{GKBA!fermions}
\index{Generalized Kadanoff-Baym Ansatz, see GKBA} (GKBA)~\cite{lipavsky_generalized_1986}
amounts to approximate  all 
interacting $G_{\blk}^{\gtrless}$  in the collision 
integral of Eq.~(\ref{eomrhot}) (including those in the 
self-energy) as 
\begin{align}
G_{\blk}^{\gtrless}(t,t')\simeq 
-G_{\blk}^{\rm R}(t,t')\,\r_{\blk}^{\gtrless}(t')+
\r_{\blk}^{\gtrless}(t)\,G_{\blk}^{\rm 
A}(t,t'),
\label{gkbae}
\end{align}
where [compare with Eq.~(\ref{G0ret})]
\begin{align}
G_{\blk}^{\rm R}(t,t')=[G_{\blk}^{\rm A}(t',t)]^{\dag}=
-i\th(t-t')
T \big\{e^{-i\int_{t'}^{t}d\bar{t}\,h_{\rm qp}(\blk,\bar{t})}\big\}.
\label{Gret}
\end{align}
This approximation to $G_{\blk}^{\rm R}$ 
corresponds to approximate $\S^{\rm R}_{\blk}(t,t')\simeq 
\d(t-t')\S_{\blk}^{\d}(t)$.
The GKBA is exact in the Hartree-Fock (HF) approximation and it is
expected to be accurate when the average time between two
consecutive collisions is longer than the quasi-particle decay
time. For systems of only electrons the GKBA allows for closing the 
equation of motion Eq.~(\ref{eomrhot}) for any diagrammatic 
approximation to $\S^{\lessgtr}$. The GKBA equation of motion 
has been successfully applied in a large variety of 
physical situations. These include 
the nonequilibrium 
dynamics~\cite{hermanns_hubbard_2014} and many-body 
localization~\cite{lev_dynamics_2014} of Hubbard 
clusters, time-dependent quantum 
transport~\cite{latini_charge_2014,tuovinen_electronic_2021},
equilibrium absorption of sodium clusters~\cite{pal_optical_2011},
real-time dynamics of the Auger decay~\cite{covito_real-time_2018},
transient absorption~\cite{perfetto_first-principles_2015,sangalli_nonequilibrium_2016,pogna_photo-induced_2016} and  
carrier 
dynamics~\cite{banyai_ultrafast_1998,vu_signature_2000,vu_time-dependent_2000,sangalli_ultra-fast_2015} of semiconductors, 
excitonic insulators out of 
equilibrium~\cite{tuovinen_comparing_2020} as well as charge 
transfer~\cite{bostrom_charge_2018} and charge 
migration~\cite{perfetto_ultrafast_2018,perfetto_first-principles_2019,perfetto_ultrafast_2020,mansson_real-time_2021} in molecular 
systems.

The GKBA is not the only ansatz to transform $G^{\lessgtr}$ 
into a functional of $\r^{<}$. An equally simple 
and legitimate ansatz can be obtained by observing that the 
group property of $G_{\blk}^{\rm R/A}$ in Eq.~(\ref{Gret}) implies 
\begin{subequations}
\begin{align}
G_{\blk}^{\rm A}(0,t')&=i G_{\blk}^{\rm A}(0,t)
G_{\blk}^{\rm R}(t,t')\quad\quad \forall t>t',
\\
G_{\blk}^{\rm R}(t,0)&=-i G_{\blk}^{\rm A}(t,t')
G_{\blk}^{\rm R}(t',0)\quad\quad \forall t<t'.
\end{align}
\end{subequations}
By the same arguments that lead to Eq.~(\ref{gkbae}) 
we then obtain 
\begin{align}
G_{\blk}^{\gtrless}(t,t')\simeq 
-\r^{\gtrless}_{\blk}(t)G_{\blk}^{\rm R}(t,t')+
\,G_{\blk}^{\rm 
A}(t,t')\r_{\blk}^{\gtrless}(t'),
\label{sgkbae}
\end{align}
which we refer to as the {\em mirrored GKBA} (MGKBA). 
Like the GKBA also the MGKBA is exact at the HF level.
In MGKBA the one-particle density matrix is on the left 
(right) of 
$G_{\blk}^{\rm R}$ ($G_{\blk}^{\rm A}$)
and it is calculated at time $t$ ($t'$). Thus, the MGKBA equation of 
motion Eq.~(\ref{eomrhot}) can be written as
\begin{align}
\frac{d}{dt}\r_{\blk}^{<}(t)+i \Big[h_{\rm 
qp}(\blk,t),\r_{\blk}^{<}(t)\Big]
&=-\G^{\rm el,>}_{\blk}(t)\r_{\blk}^{<}(t)-\G_{\blk}^{\rm el,<}(t)
\r_{\blk}^{>}(t)
\nn\\
&+{\rm h.c.},
\label{eomrhotsgkba}
\end{align}
where
\begin{align}
\G_{\blk}^{\rm el,\gtrless}(t)=\pm\int^{t} dt'\,
\S^{\gtrless}_{\blk}(t,t')G^{\rm A}_{\blk}(t',t)
\end{align}
can be interpreted as electronic scattering rates. 

An important feature of both GKBA and MGKBA is that the exact relation
\begin{align}
G_{\blk}^{>}(t,t')-G_{\blk}^{<}(t,t')=
G_{\blk}^{\rm R}(t,t')-G_{\blk}^{\rm A}(t,t')
\end{align}
is fulfilled independently of the choice of $G_{\blk}^{\rm R/A}$; 
this is a direct consequence of Eq.~(\ref{commreln<>}).

\subsection{Phonons}
\label{bosgkbasec}

The (M)GKBA for the electronic GF alone does not 
help in problems with electrons and phonons. 
The reason is that the electronic 
and phononic self-energies are functionals of $G$ and $D$ at 
different times. 
In order to close the equations of motion Eqs.~(\ref{eomrhot}) and 
(\ref{eomd<tt}) we need a (M)GKBA for phonons. 
This has been recently proposed by Karlsson et al.~\cite{karlsson_fast_2021}. 
The idea is again to 
consider the noninteracting form of the phononic 
GF~\cite{stefanucci_in-and-out_2023} 
\begin{align}
D^{<}_{0,\blq}(t,t')=D^{\rm R}_{0,\blq}(t,0)\callA
D^{<}_{0,\blq}(0,0)\callA D^{\rm A}_{0,\blq}(0,t'),
\label{bgkbastep1}
\end{align}
and then use the group property of the 
noninteracting retarded/advanced GF
\begin{align}
D^{\rm R}_{0,\blq}(t,t')=[D^{\rm A}_{0,\blq}(t',t)]^{\dag}=-i\th(t-t')\callA	
\callW_{0,\blq}(t)\callW_{0,\blq}^{-1}(t'),
\label{drafornib0}
\end{align}
where $\callW_{0,\blq}(t)=T\exp\big[-i\int_{0}^{t}d\bar{t}\;
Q(\blq)\callA\big]$. Taking into account that $\callA^{2}=1$,
for any $t>t'>0$ Eq.~(\ref{drafornib0}) implies
\begin{align}
D^{\rm R}_{0,\blq}(t,0)&=
-i\callA \callW_{0,\blq}(t)
\nn\\
&=
-i\callA\callW_{0,\blq}(t)\callW^{-1}_{0,\blq}(t')(i\callA)(-i\callA)
\callW_{0,\blq}(t')
\nn\\
&=i D^{\rm R}_{0,\blq}(t,t')\callA D^{\rm R}_{0,\blq}(t',0),
\label{gpfdgf}
\end{align}
and the like for the advanced component. 
Following the same steps leading to the electronic GKBA in 
Eq.~(\ref{gkbastep1}) we can rewrite 
Eq.~(\ref{bgkbastep1}) as 
\begin{align}	
D_{0,\blq}^{<}(t,t')= D_{0,\blq}^{\rm R}(t,t')\callA
\g_{0,\blq}^{<}(t')-
\g_{0,\blq}^{<}(t)\callA D_{0,\blq}^{\rm A}(t,t'),
\label{d<>fornib2}
\end{align}
where $\g_{0,\blq}^{<}(t)=i D_{0,\blq}^{<}(t,t)$.
An identical relation holds for $D_{0,\blq}^{>}(t,t')$ 
provided that we replace $\g_{0,\blq}^{<}$ with 
$\g_{0,\blq}^{>}$.
The GKBA for phonons~\cite{karlsson_fast_2021}
amounts to approximate  all 
interacting $D^{\gtrless}$  in the r.h.s. of Eq.~(\ref{eomd<tt}) 
(including those in the self-energy) as 
\begin{align}	
D_{\blq}^{\gtrless}(t,t')\simeq 
D_{\blq}^{\rm R}(t,t')\callA
\hat{\g}^{\gtrless}(t')-
\hat{\g}^{\gtrless}(t)\callA D_{\blq}^{\rm A}(t,t'),
\label{gkbabos}
\end{align}
where [compare with Eq.~(\ref{drafornib0})]
\begin{align}
D^{\rm R}_{\blq}(t,t')=[D^{\rm A}_{\blq}(t',t)]^{\dag}=-i\th(t-t')\callA	
\callW_{\blq}(t)\callW_{\blq}^{-1}(t'),
\label{drafornib}
\end{align}
and $\callW_{\blq}(t)=T\exp\big[-i\int_{0}^{t}d\bar{t}\;
Q_{\rm qp}(\blq,\bar{t})\callA\big]$. This approximation to $D^{\rm 
R}_{\blq}$ corresponds to approximate $\P^{\rm R}_{\blq}(t,t')\simeq 
\d(t-t')\P_{\blq}^{\d}(t)$. 

Like in the electronic case the phononic GKBA is not the 
only ansatz to transform $D^{\lessgtr}$ into a functional 
of $\g^{<}$. By definition 
\begin{subequations}
\begin{align}
D^{\rm A}_{\blq}(0,t')&=D^{\rm A}_{\blq}(0,t)(i\callA)
D^{\rm R}_{\blq}(t,t')\quad\quad \forall t>t',
\\
D^{\rm R}_{\blq}(t,0)&=-D^{\rm A}_{\blq}(t,t')
(i\callA)D^{\rm R}_{\blq}(t',0)\quad\quad \forall t<t'.
\end{align}
\end{subequations}
By the same arguments leading to Eq.~(\ref{gkbabos})
we then find an equally 
simple and legitimate ansatz, which we refer to as the MGKBA for phonons: 
\begin{align}	
D^{\gtrless}_{\blq}(t,t')\simeq 
\g^{\gtrless}(t)\callA D_{\blq}^{\rm R}(t,t')-
D_{\blq}^{\rm A}(t,t')\callA\g^{\gtrless}(t').
\label{sgkbabos}
\end{align}
In MGKBA the equation of motion Eq.~(\ref{eomd<tt}) can be written as
\begin{align}
\frac{d}{dt}\g^{<}_{\blq}(t)&+i
\Big(\callA Q_{\rm qp}(\blq,t)\g^{<}_{\blq}(t)-
\g^{<}_{\blq}(t)Q_{\rm qp}(\blq,t)\callA\Big)
\nn\\
&=-\G^{\rm ph,>}_{\blq}(t)\g^{<}_{\blq}(t)+
\G^{\rm ph,<}_{\blq}(t)\g^{>}_{\blq}(t)+{\rm h.c.},
\label{eomd<ttrate}
\end{align}
where 
\begin{align}
\G^{\rm ph,\lessgtr}_{\blq}(t)=\callA\int^{t} dt'\,\P^{\lessgtr}_{\blq}(t,t')
D_{\blq}^{\rm A}(t',t)\callA
\end{align}
can be interpreted as phononic scattering rates.

For phonons a physically sensible approximation to $\P^{\d}$ is the 
clamped-nuclei plus static approximation~\cite{stefanucci_in-and-out_2023}, i.e., 
\begin{align}
\P^{ij,\d}_{\blq\a\a'}=\d_{i1}\d_{j1} 
\sum_{\subalign{&\blk,\blk'\\&\m\m'\\ &\n\n'}}
g_{\blq\a,\m\n}(\blk)\chi^{\rm R}_{\blk,\blk',\subalign{&\n\n'\\ &\m\m'}}
(\blq;\w=0)g^{\ast}_{\blq\a',\m'\n'}(\blk'),
\label{pideltamin}
\end{align}
where $\chi$ is the 
response function at clamped nuclei (for the index structure see 
Fig.~\ref{SigmaGWFM}). 
Such self-energy renormalizes the 
block $(1,1)$ of the matrix $Q$, see Eq.~(\ref{qpQ}), 
which in the basis of the BO normal 
modes becomes diagonal, see Eq.~(\ref{exactid}). Thus the whole matrix 
$Q_{\rm qp,\a\a'}(\blq)=\d_{\a\a'}\left(\begin{array}{cc}
\w_{\blq\a}^{2} & 0 
\\
0 & 1
\end{array}\right)$ becomes diagonal and the retarded GF simplifies to
with
\begin{align}
D^{\rm R}_{\blq\a\a'}(t,t')=i\d_{\a\a'}\frac{ \th(t-t')}{2\w_{\blq\a}}&\left[
e^{i\w_{\blq\a}(t-t')}\left(\begin{array}{cc}\!\!1 & \!\!-i \w_{\blq\a} 
\\ \!\!i \w_{\blq\a} &  \!\!\w_{\blq\a}^{2}\end{array}
\!\!\right)\right.
\nn\\
-&
\left.e^{-i\w_{\blq\a}(t-t')}\left(\begin{array}{cc}\!\!1 & \!\!i 
\w_{\blq\a} 
\\ \!\!-i \w_{\blq\a} &  \!\!\w_{\blq\a}^{2}\end{array}
\!\!\right)
\right].
\label{DRc+sBO}
\end{align}
This approximated form depends only on the time difference and can be 
Fourier transformed. It is easy to verify that 
$D^{11,\rm R}_{\blq\a}(\w)=1/[(\w+\iu\eta)^{2}-\w_{\blq\a}^{2}]$.
We further note that also for 
phonons the exact property 
\begin{align}
D_{\blq}^{>}(t,t')-D_{\blq}^{<}(t,t')=
D_{\blq}^{\rm R}(t,t')-D_{\blq}^{\rm A}(t,t')
\end{align}
is fulfilled in both GKBA and MGKBA regardless of the choice of $D_{\blq}^{\rm 
R/A}$; this is a direct consequence of Eq.~(\ref{g>=g<+a}) and 
$\callA^{2}=1$.

\section{Electronic and phononic self-energies}
\label{sesec}

Through the (M)GKBA for electrons and phonons the equations of motion 
(\ref{eomrhot}) and (\ref{eomd<tt}) become integro-differential 
equations for $\r^{<}$ and 
$\g^{<}$ for any diagrammatic approximation to the 
self-energies $\S$ and $\P$. 
As a general remark we observe that the treatment of the Keldysh 
components of the self-energies lacks consistency in (M)GKBA. The 
retarded/advanced self-energies are usually evaluated in some 
quasi-particle approximation (hence they are time-local)
and have the only purpose of
improving the retarded/advanced GFs in the ansatzes.
The lesser/greater self-energies do instead keep their diagrammatic 
form, see below. 

\begin{figure}[tbp]
    \centering
\includegraphics[width=0.49\textwidth]{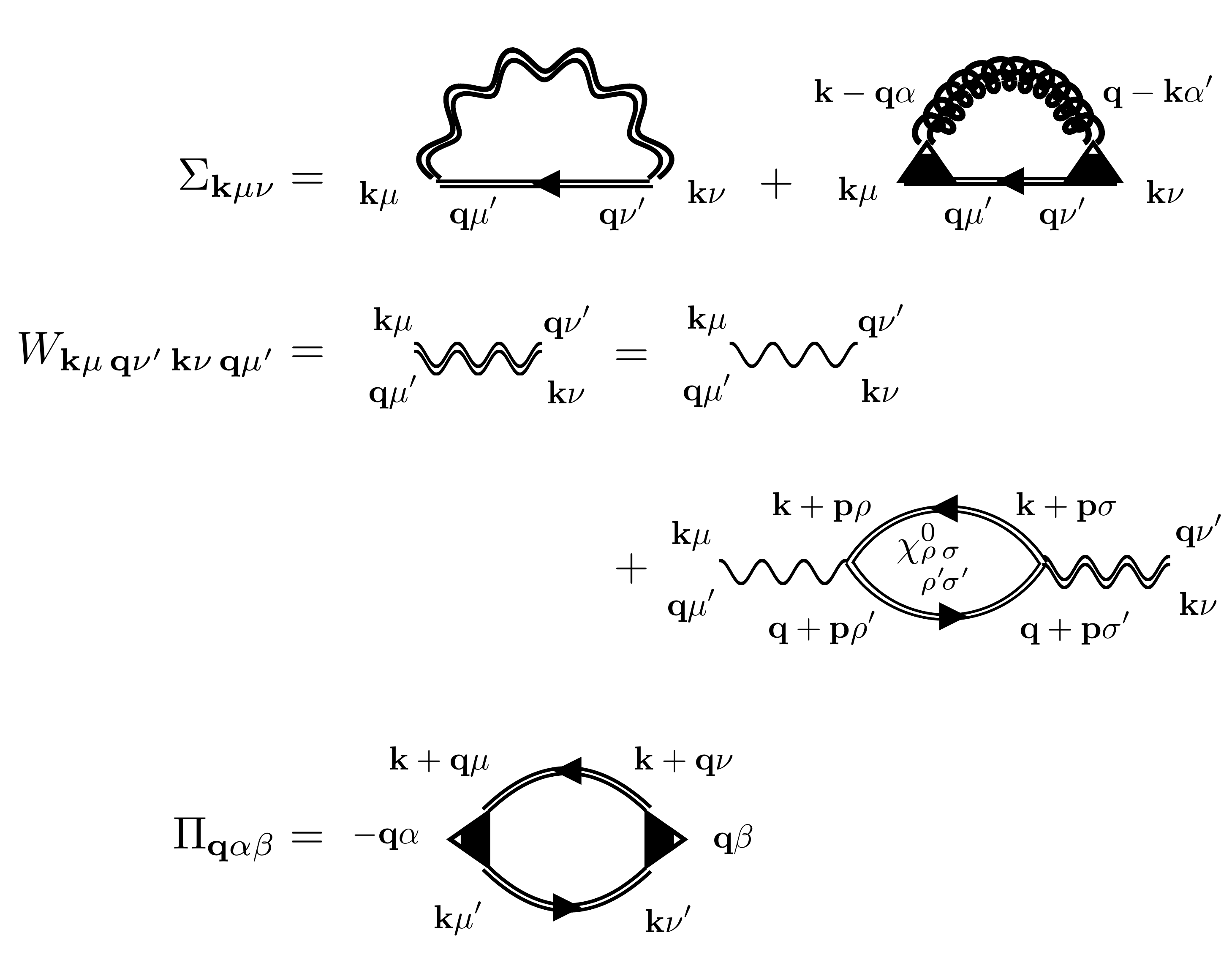}
\caption{(Top) Electronic self-energy in the GW plus Fan-Migdal 
approximation. (Middle) RPA equation for the screened interaction 
$W$. (Bottom) Phononic self-energy.}
\label{SigmaGWFM}
\end{figure}

We here consider the many-body approximation to  
$\S$ and $\P$ which, as we see in Section~\ref{recoverysec}, enables us to recover the 
Boltzmann equations. The $GW$ plus Fan-Migdal approximation for $\S$ 
and the bare bubble approximation for $\P$ are illustrated in 
Fig.~\ref{SigmaGWFM}. In labelling  the internal vertices we  
already take into account  the conservation of the quasi-momentum. 

For the $GW$ self-energy 
$\S^{{\rm GW},\gtrless}=i G^{\gtrless}W^{\lessgtr}$ we need to express
 $W^{\lessgtr}$ in terms of $G^{<}$ and $G^{>}$. 
Starting from the Dyson equation $W=v+vPW$ one easily find 
$W^{\gtrless}=W^{\rm R}P^{\gtrless}W^{\rm A}$.
To reduce the complexity of the  equations we take $P=\chi^{0}=-i GG$ 
(RPA approximation) and 
evaluate $W^{\rm R/A}$ 
in the statically screened approximation. We  
have~\cite{haug_quantum_1994,marini_competition_2013}
\begin{align}
\S^{{\rm GW},<}_{\blk\m\n}(t,t')&=i\sum_{\blq\n'\m'}
W^{<}_{\blk\m\,\blq\n'\,\blk\n\,\blq\m'}(t,t')
G^{<}_{\blq\m'\n'}(t,t')
\nn\\
&=\sum_{\blq\n'\m'}\sum_{\blp\r\s\r'\s'}
W_{\blk\m\,\blq+\blp\r'\,\blk+\blp\r\,\blq\m'}G^{>}_{\blq+\blp\s'\r'}(t',t)
\nn\\
&\times
G^{<}_{\blk+\blp\r\s}(t,t')W_{\blk+\blp\s\,\blq\n'\,\blk\n\,\blq+\blp\s'}
G^{<}_{\blq\m'\n'}(t,t'),
\label{GWstatscr}
\end{align}
The greater component of the self-energy is obtained by exchanging 
$>\;\leftrightarrow\; <$. We can add to the $GW$ self-energy the 
second-order exchange diagram with statically screened $W$ 
lines~\cite{steinhoff_nonequilibrium_2016}. This amounts to 
replace 
$W_{\blk_{1}\m_{1}\,\blk_{2}\m_{2}\,\blk_{3}\m_{3}\,\blk_{4}\m_{4}}
\to \frac{1}{2}[W_{\blk_{1}\m_{1}\,\blk_{2}\m_{2}\,\blk_{3}\m_{3}\,\blk_{4}\m_{4}}
-W_{\blk_{1}\m_{1}\,\blk_{2}\m_{2}\,\blk_{4}\m_{4}\,\blk_{3}\m_{3}}]$ 
in Eq.~(\ref{GWstatscr}). 

Next we consider the Fan-Migdal self-energy in Fig.~\ref{SigmaGWFM}. 
We take a bare $e$-$ph$ coupling $g$ and later discuss how to 
dress it. We have
\begin{align}
\S^{{\rm 
FM},<}_{\blk\m\n}(t,t')&=i \sum_{\blq\n'\m'\a\a'}
g_{\blq-\blk\a,\m\m'}(\blk)
D_{\blk-\blq\a\a'}^{11,<}(t,t')
\nn\\
&\times G^{<}_{\blq\m'\n'}(t,t')g^{\ast}_{\blq-\blk\a',\n\n'}(\blk),
\label{fmsebg}
\end{align}
where we use the property in Eq.~(\ref{ephcoupprop}).
The Fan-Migdal self-energy calculated with a screened $e$-$ph$ 
coupling $g^{d}=(1+WP)g=(1+v\chi)g$
is more involved. 
To account for screening effects to some degree we can make the 
clamped-nuclei plus static approximation used to estimate $\P^{\d}$, 
i.e., $g^{d}=[1+v\chi(\w=0)]g$. In this way the screened $e$-$ph$ 
coupling is still time-local and the mathematical form of the 
Fan-Migdal self-energy is identical to Eq.~(\ref{fmsebg}) with 
$g\to g^{d}$. The greater component is obtained by exchanging 
$>\;\leftrightarrow\; <$.

We finally analyze the phononic self-energy. Keeping an eye on 
Fig.~\ref{SigmaGWFM} we have
\begin{align}
\P^{ii',<}_{\blq\a\b}(t,t')&=-i\d_{i1}\d_{i'1} \sum_{\blk\m\n\m'\n'}
g_{\blq\a,\m'\m}(\blk)
\nn\\
&\times
G^{<}_{\blq+\blk\m\n}(t,t')G^{>}_{\blk\n'\m'}(t',t)g^{\ast}_{\blq\b,\n'\n}(\blk),
\label{Piq}
\end{align}
where in the second equality we use Eq.~(\ref{ephcoupprop}). 
The greater component is obtained by exchanging 
$>\;\leftrightarrow\; <$.
Dressing the $e$-$ph$ coupling is here less straightforward since
only one $g$ should be 
dressed~\cite{calandra_adiabatic_2010,feliciano_electron-phonon_2017,marini_equilibrium_2023,berges_phonon_2023}.
Implementing the 
clamped-nuclei plus static approximation to only one $g$  would lead 
to the violation of the hermiticity properties of the self-energy. 
No violation occurs  if 
the {\em nonlocal in time} dressed coupling is used. 
However, the equations of motion become more complex and, to the best of 
our knowledge, no efforts have been made to solve them thus far. 
Currently, all nonequilibrium state-of-the-art 
methods dress {\em both} $g$'s in  
Eq.~(\ref{Piq}), thereby suffering of a double counting problem. 
The SEPE do not resolve this issue either.

Through the GKBA for electrons and phonons the self-energies  
$\S^{\rm GW,\lessgtr}(t,t')$, $\S^{\rm FM,\lessgtr}(t,t')$ and 
$\P^{\lessgtr}(t,t')$ with $t>t'$
become functionals of $\r^{<}(t')$ and $\g^{<}(t')$ with $t'<t$. 
Therefore the GKBA equations (\ref{eomrhot}) 
and (\ref{eomd<tt})  carry memory, i.e., the density 
matrices at time $t$ depend on the history. 
This scheme has been recently implemented to study the relaxation of 
electrons and phonons in a 
photoexcited MoS$_{2}$ monolayer~\cite{perfetto_real-time_2023}. 
On the contrary the MGKBA leads to equations of motion with no memory 
since the self-energies become functionals of $\r^{<}(t)$ and 
$\g^{<}(t)$.

\section{Semiconductor electron-phonon equations}
\label{sepesec}

The SEPE are derived from the MGKBA equations after a series of 
simplifications, and they apply 
to crystals with a finite gap in the one-particle spectrum, like 
semiconductors or insulators. 
In Appendix~\ref{gkbafailure}, we demonstrate that  the 
{\em same} 
simplifications in the GKBA framework give rise to unphysical 
divergences and the inability to recover the Boltzmann equations.
At least in this context, MGKBA is superior to 
GKBA.
 
The first simplification (S1) consists in using an MGKBA
with diagonal density matrices and 
diagonal retarded/advanced GF -- diagonality here 
refers to band indices for 
the electrons and mode indices for the phonons. 
For the electronic part we work in the eigenbasis 
of
the equilibrium quasi-particle Hamiltonian,
hence $h_{\rm qp,\m\m'}(\blk)=\d_{\m\m'}\e_{\blk\m}$, 
and ignore the time-dependence of $h_{\rm qp}(\blk,t)$ in 
Eq.~(\ref{Gret}). Then Eq.~(\ref{sgkbae}) becomes
$G^{\lessgtr}_{\blk\m\n}(t,t')=\d_{\m\n}G^{\lessgtr}_{\blk\m\m}(t,t')$, with
\begin{align}
G^{<}_{\blk\m\m}(t,t')=i
e^{-i \e_{\blk\m}(t-t')}
\big[
\th(t-t')f^{\rm el}_{\blk\m}(t)+\th(t'-t)f^{\rm el}_{\blk\m}(t')\big].
\label{gkbaapprox}
\end{align}
The expression for $G^{>}_{\blk\m\n}(t,t')$ is identical provided that 
$f^{\rm el}_{\blk\m}\to 
f^{\rm el}_{\blk\m}-1$, see Eq.~(\ref{commreln<>}). For the phononic part we work 
in the basis of the BO normal modes. Taking into account 
Eqs.~(\ref{gammamatele}) and (\ref{DRc+sBO}), the phononic MGKBA in 
Eq.~(\ref{sgkbabos})  yields 
$D^{\lessgtr}_{\blq\a\b}(t,t')=\d_{\a\b}D^{\lessgtr}_{\blq\a\a}(t,t')$, with
\begin{align}
D^{11,\lessgtr}_{\blq\a\a}(t,t')&=\frac{\th(t-t')}{2i\w_{\blq\a}}
\big[B^{\lessgtr\ast}_{\blq\a}(t)e^{-i\w_{\blq\a}(t-t')}+
B^{\gtrless}_{-\blq\a}(t)e^{i\w_{\blq\a}(t-t')}\big]
\nn\\
+&\frac{\th(t'-t)}{2i\w_{\blq\a}}
\big[B^{\lessgtr}_{\blq\a}(t')e^{-i\w_{\blq\a}(t-t')}+
B^{\gtrless\ast}_{-\blq\a}(t')
e^{i\w_{\blq\a}(t-t')}\big],
\label{D11<gkba}
\end{align}
\begin{align}
D^{12,\lessgtr}_{\blq\a\a}(t,t')&=\frac{\th(t-t')}{2}
\big[
B^{\lessgtr\ast}_{\blq\a}(t)e^{-i\w_{\blq\a}(t-t')}-
B^{\gtrless}_{-\blq\a}(t)e^{i\w_{\blq\a}(t-t')}\big]
\nn\\
+&\frac{\th(t'-t)}{2}
\big[ C^{\lessgtr}_{\blq\a}(t')e^{-i\w_{\blq\a}(t-t')}-
 C^{\gtrless\ast}_{-\blq\a}(t')e^{i\w_{\blq\a}(t-t')}\big],
\label{D12<gkba}
\end{align}
and 
\begin{subequations}
\begin{align}
B^{<}_{\blq\a}(t)&=f^{\rm ph}_{\blq\a}(t)+\Th_{\blq\a}(t),
\label{B<occcoh}
\\
B^{>}_{\blq\a}(t)&=f^{\rm ph}_{\blq\a}(t)+1+\Th_{\blq\a}(t),
\label{B>occcoh}
\\
C^{<}_{\blq\a}(t)&=f^{\rm 
ph}_{\blq\a}(t)-\Th_{\blq\a}(t),
\label{C<occcoh}
\\
C^{>}_{\blq\a}(t)&=f^{\rm 
ph}_{\blq\a}(t)+1-\Th_{\blq\a}(t).
\label{C>occcoh}
\end{align}
\label{BClikecomb}
\end{subequations}
As we see later, the GF $D^{22}$ is not needed. 

In the following we derive the SEPE for the electronic occupations 
and polarizations, nuclear displacements, and phononic occupations 
and coherences. 
Without any loss of generality, we assume that the 
system is initially in equilibrium at negative times and it is 
subsequently perturbed by a driving field of finite duration 
$T_{\rm drive}$; hence 
$\blA(\blr,t)=0$ for $t<0$ and for $t>T_{\rm drive}$.

\subsection{Electronic occupations and polarizations}
The second simplification (S2) to derive the SEPE 
consists in approximating 
\begin{align}
h_{\m\n}(\blk)+\S^{\d}_{\blk\m\n}(t)\simeq 
\d_{\m\n}\tilde{\e}_{\blk\m}(t),
\end{align}
where
\begin{align}
\tilde{\e}_{\blk\m}(t)=\e_{\blk\m}
+\sum_{\blk'\m'}
\big(v_{\blk\m\blk'\m'\blk'\m'\blk\m}-W_{\blk\m\blk'\m'\blk\m\blk'\m'}\big)
\D f^{\rm el}_{\blk'\m'}(t),
\label{sphamsbe}
\end{align}
and 
\begin{align}
\D f^{\rm el}_{\blk\m}(t)\equiv f^{\rm el}_{\blk\m}(t)-f^{\rm 
el}_{\blk\m}(0).
\end{align}
The second term in Eq.~(\ref{sphamsbe}) 
is the change in the diagonal part 
of HSEX
potential due to a change of the occupations. At time $t=0$ the 
Rabi frequencies and the nuclear displacements vanish, and therefore
$h_{\rm 
qp,\m\m'}(\blk,0)=h_{\m\m'}(\blk)+\S^{\d}_{\blk\m\m'}(0)=\d_{\m\m'}\e_{\blk\m}$, 
as it should.

Let $S^{\rm el}_{\blk}(t)$ be the r.h.s. of 
Eq.~(\ref{eomrhot}); we call this quantity the {\em electronic 
scattering term}. We then have
\begin{align}
\frac{d}{dt}f^{\rm el}_{\blk\m}+i\sum_{\n\neq \m}
\Big(\W^{\rm ren}_{\blk\m\n}\,p_{\blk\n\m}-p_{\blk\m\n}\,\W^{\rm 
ren}_{\blk\n\m}\Big)=
S^{\rm el}_{\blk\m\m},
\label{sbeocc}
\end{align}
\begin{align}
\frac{d}{dt}p_{\blk\m\n}&+i\big(\tilde{\e}_{\blk\m}-\tilde{\e}_{\blk\n}\big)p_{\blk\m\n}
+i\W^{\rm ren}_{\blk\m\n}\big(f^{\rm el}_{\blk\n}-f^{\rm el}_{\blk\m}\big)
\nn\\&+i
\sum_{\n'\neq \n}
\W^{\rm ren}_{\blk\m\n'}\,p_{\blk\n'\n}-i\sum_{\n'\neq 
\m}p_{\blk\m\n'}\,\W^{\rm ren}_{\blk\n'\n}=
S^{\rm el}_{\blk\m\n},
\label{sbepol}
\end{align}
where we define the renormalized Rabi frequencies
\begin{align}
\W^{\rm ren}_{\blk\m\n}(t)\equiv \W_{\blk\m\n}(t)+
\sum_{\a}g_{{\mathbf 0}\a,\m\n}(\blk)
U_{{\mathbf 0}\a}(t).
\label{renrabi}
\end{align}
The equations of motion Eqs.~(\ref{sbeocc}) and (\ref{sbepol}) 
with 
$S^{\rm el}=0$  are 
equivalent to solving the Bethe-Salpeter 
equation~\cite{attaccalite_real-time_2011,svl-book}; they
have been recently implemented to investigate the 
dynamics of coherent 
excitons~\cite{perfetto_pump-driven_2019,sangalli_excitons_2021,chan_giant_2023}.
The diagonal part of the scattering term is the sum of the $GW$ and Fan-Migdal 
contributions:
\begin{align}
S^{\rm el}_{\blk\m\m}(t)=S^{\rm GW}_{\blk\m\m}(t)
+S^{\rm FM}_{\blk\m\m}(t).
\label{scattel}
\end{align}
By using the simplified MGKBA of 
Eqs.~(\ref{gkbaapprox}) and (\ref{D11<gkba}) when
calculating the self-energies in Eqs.~(\ref{GWstatscr}) and 
(\ref{fmsebg}), 
both contributions in Eq.~(\ref{scattel}) 
can be expressed in terms of $f^{\rm 
el}$, $f^{\rm ph}$ and $\Th$. The third simplification (S3) of the SEPE 
is the Markov approximation for all exponential integrals, i.e., 
$\int_{0}^{t}e^{i Et}\simeq \p\d(E)$. After some 
straightforward algebra we find
\begin{widetext}
\begin{align}
S^{\rm GW}_{\blk\m\m}&=2\p
\sum_{\blq\m'}\sum_{\blp\n\n'}
\big|W_{\blk\m\,\blq+\blp\n'\,\blk+\blp\n\,\blq\m'}\big|^{2}
\d\big(\e_{\blk+\blp\n}+
\e_{\blq\m'}-\e_{\blq+\blp\n'}
-\e_{\blk\m}\big)
\nn
\\&\times
\Big[\big(f^{\rm el}_{\blq+\blp\n'}-1\big)\big(f^{\rm el}_{\blk\m}-1\big)
f^{\rm el}_{\blk+\blp\n}f^{\rm el}_{\blq\m'}
-
f^{\rm el}_{\blq+\blp\n'}f^{\rm el}_{\blk\m}
\big(f^{\rm el}_{\blk+\blp\n}-1\big)
\big(f^{\rm el}_{\blq\m'}-1\big)\Big],
\label{sbescattdiagmarkov}	
\end{align}
and
\begin{align}
S^{\rm FM}_{\blk\m\m}&=2\p
\sum_{\blq\n\a}\frac{\big|g_{\blq-\blk\a,\m\n}(\blk)\big|^{2}}{\w_{\blk-\blq\a}}
\Big\{\d\big(
\e_{\blq\n}-\e_{\blk\m}
+\w_{\blk-\blq\a}
\big)
\Big[
\big(f^{\rm el}_{\blq\n}-1\big)f^{\rm el}_{\blk\m}\Re\big[B^{>}_{\blk-\blq\a}\big]
-f^{\rm el}_{\blq\n}\big(f^{\rm el}_{\blk\m}-1\big)\Re\big[B^{<}_{\blk-\blq\a}\big]
\Big]
\nn\\
&+\d\big(
\e_{\blq\n}-\e_{\blk\m}
-\w_{\blk-\blq\a}
\big)
\Big[
\big(f^{\rm el}_{\blq\n}-1\big)f^{\rm el}_{\blk\m}\Re\big[B^{<}_{\blq-\blk\a}\big]
-
f^{\rm el}_{\blq\n}\big(f^{\rm el}_{\blk\m}-1\big)\Re\big[B^{>}_{\blq-\blk\a}\big]
\Big]\Big\}.
\label{sbescattdiagFMgenmarkov}
\end{align}
\end{widetext}
The Fan-Migdal scattering term depends on both phononic occupations 
and coherences. It is easy to verify that $\sum_{\blk\m}S^{\rm 
GW}_{\blk\m\m}(t)=\sum_{\blk\m}S^{\rm 
FM}_{\blk\m\m}(t)=0$, which guarantees the conservation of the total 
number of electrons.
We mention that the GKBA gives identical 
results for both scattering terms.

The off-diagonal scattering term in Eq.~(\ref{sbepol}) 
is usually simplified as 
\begin{align}
S^{\rm el}_{\blk\m\n}(t)\simeq -
\G^{\rm pol}_{\blk\m\n}(t) p_{\blk\m\n}(t),\quad\quad \forall \m\neq 
\n\,.
\label{sbescattpol}
\end{align}
In fact, the polarizations carry information on the electronic 
coherence and are expected to vanish after the 
photo-excitation.   
The polarization rates $\G^{\rm pol}_{\blk\m\n}$ 
can be calculated as outlined in 
Refs.~\cite{marini_ab-initio_2008,chan_exciton_2023}, although they
are often treated as fitting 
parameters. A semi-empirical way to 
estimate them is based on the observation that 
the electronic scattering term in Eq.~(\ref{scattel}) has 
the following mathematical structure
\begin{align}
S^{\rm el}_{\blk\m\m}(t)=-2\G^{\rm el,>}_{\blk\m\m}(t)f_{\blk\m}(t)
-2\G^{\rm el,<}_{\blk\m\m}(t)\big(f_{\blk\m}(t)-1\big).
\label{scatteldiag}
\end{align}
It is easy to verify that $\G^{\rm el,\lessgtr}_{\blk\m\m}(t)\geq 0$ for 
vanishing phononic coherences, i.e., $\Th_{\blq\a}=0$.
Comparing Eq.~(\ref{scatteldiag}) 
with Eq.~(\ref{eomrhotsgkba}) we infer that our simplifications 
have led to diagonal electronic scattering rates.
Taking the $(\m,\n)$ element of Eq.~(\ref{eomrhotsgkba}) 
we then obtain the following expression for the polarization rates
\begin{align}
\G^{\rm pol}_{\blk\m\n}(t)=\G^{\rm el,>}_{\blk\m\m}(t)+\G^{\rm el,<}_{\blk\m\m}(t)
+\G^{\rm el,>}_{\blk\n\n}(t)+\G^{\rm el,<}_{\blk\n\n}(t).
\label{polscattrate}
\end{align}
It is worth remarking that the Markovian approximation of the GKBA 
equations of motion  leads to unphysical polarization rates, see 
Appendix~\ref{gkbafailure}. 

For a full time-dependent framework of electrons and phonons
Eqs.~(\ref{sbeocc}) and (\ref{sbepol}) must be coupled to 
the equations 
of motion for the nuclear displacements  and phononic density matrix.
The treatment of phonons 
necessitates a preliminary discussion on the equilibrium response 
function at clamped nuclei. We here consider the RPA  
$\chi=\chi^{0}+\chi^{0}v\chi=\chi^{0}+\chi^{0}W\chi^{0}$. 
Omitting time integrals and 
momentum labels, and using for $\chi$ the same index structure as in 
Fig.~\ref{SigmaGWFM}  we have  
\begin{align}
\chi_{\subalign{&\m\m'\\ &\n\n'}}=\d_{\m\m'}\d_{\n\n'}
\chi^{0}_{\subalign{&\m\\ &\n}}+\chi^{0}_{\subalign{&\m\\ &\n}}
W_{\m\n'\m'\n}\chi^{0}_{\subalign{&\m'\\ &\n'}},
\label{rpachi}
\end{align}
where we take into account that $\chi^{0}_{\subalign{&\m\m'\\ &\n\n'}}
=\d_{\m\m'}\d_{\n\n'}\chi^{0}_{\subalign{&\m\\ &\n}}$ due to 
simplification (S1). For a semiconductor at low temperature 
$\chi^{0,\lessgtr}_{\subalign{&\m\\ &\n}}\simeq 0$ if $\m$ and $\n$ are both 
conduction or valence bands. 
Therefore, the only 
sizable elements of the response function
are those for which the indices of 
the pairs $(\m,\m')$ 
and $(\n,\n')$ are either conduction-valence or valence-conduction.

\subsection{Nuclear displacements}

The equation of motion Eq.~(\ref{sducp}) 
for the nuclear displacements contains 
the elastic tensor $K$.
It would be desirable to formulate a simplified equation where 
the elastic tensor is renormalized by the phononic self-energy, 
giving rise to the BO frequencies, see Eq.~(\ref{exactid}). 
From Eqs.~(\ref{sbeocc}) and 
(\ref{sbepol}) we infer that 
only the polarizations $p_{\blk\m\n}$ with indices 
$(\m,\n)$ either conduction-valence or valence-conduction 
change linearly with the  
driving field, since it is only in this case that
$f^{\rm el}_{\blk\n}(0)-f^{\rm el}_{\blk\m}(0)$ is sizable. 
All other elements of the 
electronic density matrix
change at least quadratically.
We then write the polarizations as
\begin{align}
p_{\blk\m\n}=p^{\rm inter}_{\blk\m\n}+
p^{\rm intra}_{\blk\m\n},
\end{align}
where $p^{\rm inter}_{\blk\m\n}$ is non vanishing only if $\m$ is a valence 
(conduction) band and $\n$ is a conduction (valence) band whereas 
$p^{\rm intra}_{\blk\m\n}$ is nonvanishing only if $\m$ and $\n$ are both valence 
or conduction bands. Taking into account that $p_{\blk\m\n}(0)=0$
we can rewrite the first term in 
the r.h.s. of Eq.~(\ref{sducp}) as 
\begin{align}
\sum_{\blk\m\n}g_{{\mathbf 
0}\a,\n\m}(\blk)\D \r^{<}_{\blk\m\n}&=
\sum_{\blk\m\n}g_{{\mathbf 
0}\a,\n\m}(\blk)p^{\rm inter}_{\blk\m\n}
\nn\\&+
\sum_{\blk\m\n}g_{{\mathbf 
0}\a,\n\m}(\blk)\big[\d_{\m\n}\D f^{\rm el}_{\blk\m}+
p^{\rm intra}_{\blk\m\n}\big].
\label{gdeltan}
\end{align}
The fourth simplification (S4) consists in expressing $p^{\rm 
inter}_{\blk\m\n}$ using the Kubo formula, 
and in treating the nuclei in the 
Ehrenfest approximation.
Omitting the dependence on momenta we have, see 
Appendix~\ref{ehchisec},
\begin{align}
p^{\rm inter}_{\m\n}(t)=\!\int \!\!dt'\!\sum_{\m'\n'}\chi^{\rm 
R}_{\subalign{&\m\m'\\\ 
&\n\n'}}(t,t')\Big[\W_{\m'\n'}(t')+\sum_{\a}g_{\a,\m'\n'}
U_{\a}(t')\Big],
\label{lrfehdyn}
\end{align}
with $\chi$ the response function in the clamped-nuclei approximation.
For a response function that decays fast as $|t-t'|\to \iif$ we can 
evaluate the slowly varying function $U_{\a}$
at time $t$ instead of $t'$, and hence perform the time integral
$\int \!\!dt'\chi^{\rm R}(t,t')=\chi^{\rm R}(\w=0)$. Taking into 
account Eq.~(\ref{rpachi}) and the discussion below it, we then 
find the following 
important result
\begin{align}
\sum_{\m\n}g_{\a,\n\m}\,p^{\rm inter}_{\m\n}
&+\sum_{\a'}K_{\a\a'}U_{\a'}	
=\w^{2}_{\a}U_{\a}
\nn\\
&+\int dt'\!\sum_{\subalign{&\m\m'\\\ 
&\n\n'}}g_{\a,\n\m}\chi^{\rm 
R}_{\subalign{&\m\m'\\\ 
&\n\n'}}(t,t')\W_{\m'\n'}(t'),
\end{align}
where we have recognized the phononic self-energy in 
the clamped-nuclei plus static approximation, see Eq.~(\ref{pideltamin}), and 
used Eq.~(\ref{exactid}). In conclusion, the 
equation of motion Eq.~(\ref{sducp}) becomes 
(reintroducing the dependence on momenta)
\begin{align}
\frac{d^{2}}{dt^{2}}U_{{\mathbf 0}\a}(t)
&+\w^{2}_{{\mathbf 0}\a}U_{{\mathbf 0}\a}(t)
\nn\\
&=-\sum_{\blk\m\n}g_{{\mathbf 
0}\a,\n\m}(\blk)\big[\d_{\m\n}\D f^{\rm el}_{\blk\m}(t)+
p^{\rm intra}_{\blk\m\n}(t)\big]
\nn\\
&-\int dt'\sum_{\subalign{&\blk\blk'\\&\m\m'\\ 
&\n\n'}}g_{{\mathbf 
0}\a,\n\m}(\blk)\chi^{\rm 
R}_{\blk\blk',\subalign{&\m\m'\\\ 
&\n\n'}}({\mathbf 
0};t,t')\W_{\blk'\m'\n'}(t').
\label{sducp2}
\end{align}
We emphasize that Eq.~(\ref{sducp2})
differs from the equation of motion that follows from typical $e$-$ph$ model 
Hamiltonians, where Eq.~(\ref{h0phs}) is replaced by 
$\hat{H}_{0,ph}=\sum_{\blq\a}\w_{\blq\a}\hat{b}^{\dag}_{\blq\a}\hat{b}_{\blq\a}$. 
In this case one would find $\frac{d^{2}}{dt^{2}}U_{{\mathbf 
0}\a}+\w^{2}_{{\mathbf 0}\a}U_{{\mathbf 0}\a}(t)=
-\sum_{\blk\m\n}g_{{\mathbf 
0}\a,\n\m}(\blk)\D\r_{\blk\m\n}(t)$, which involves 
the full density matrix, not just the intra-only 
elements; moreover the last term in Eq.~(\ref{sducp2}) is missing.
As pointed out in 
Refs.~\cite{marini_many-body_2015,stefanucci_in-and-out_2023},
model Hamiltonians suffer from a double 
renormalization of the phononic frequencies.

\subsection{Phononic occupations and coherences}
\label{phdmatsubsec}

The starting point is here the 
equation of motion Eq.~(\ref{eomd<tt}) in the basis of the BO normal 
modes. Using the quasi-particle $2\times 2$ matrix
$Q_{\rm qp}$ derived above Eq.~(\ref{DRc+sBO}) we can write
\begin{align}
\frac{d}{dt}\g^{<}_{\blq\a\a}(t)&+
\left(\begin{array}{cc}
0 & -1 
\\
\w_{\blq\a}^{2} & 0
\end{array}\right)\g^{<}_{\blq\a\a}(t)-
\g^{<}_{\blq\a\a}(t)
\left(\begin{array}{cc}
0 & -\w_{\blq\a}^{2}
\\
1 & 0
\end{array}\right)
\nn\\
&=S^{\rm ph}_{\blq\a\a}(t)
\label{sepeqgamma}
\end{align}
where
\begin{align}
S^{\rm ph}_{\blq\a\a}(t)\equiv \callA\int^{t}dt'&\big[
\P^{>}_{\blq\a\a}(t,t') D^{<}_{\blq\a\a}(t',t)
\nn\\
&-\P^{<}_{\blq\a\a}(t,t') D^{>}_{\blq\a\a}(t',t)\big]+{\rm 
h.c.}
\label{phonscatt}
\end{align}
is the phononic scattering term. 
Let us inspect the elements of the $2\times 2$ matrix $S^{\rm ph}$. 
Naming the integral in Eq.~(\ref{phonscatt}) with the 
letter $J$ -- hence $J$ is a $2\times 2$ matrix --
and taking into account that the self-energy has 
only one nonvanishing element, which is the $(1,1)$, we find
\begin{align}
S^{\rm ph}&=\callA J+{\rm h.c.}=
\left(\!\begin{array}{cc} 0 & i \\ -i & 0 \end{array}\!\right)
\left(\!\begin{array}{cc} J^{11} & J^{12} \\ 0 & 0 \end{array}\!\right)
+{\rm h.c.}
\nn\\&=\left(\!\begin{array}{cc} 0 & i J^{11\ast} \\ -i J^{11} 
& -i J^{12}+i J^{12\ast} \end{array}\!\right).
\nn
\end{align}
Thus, to calculate  
$S^{\rm ph}_{\blq\a\a}$ we only need the $(1,1)$ and 
$(1,2)$ elements of the phononic GF, whose  
MGKBA form has been 
derived in Eqs.~(\ref{D11<gkba}) and (\ref{D12<gkba}).
Evaluating the phononic self-energy, see Eq.~(\ref{Piq}), at the 
MGKBA GFs, and implementing the Markov approximation (S3)
we obtain
\begin{widetext}
\begin{align}
S^{21,\rm ph}_{\blq\a\a}&=-i\p\sum_{\blk\m\n}
\frac{|g_{\blq\a,\m\n}(\blk)|^{2}}{2\w_{\blq\a}}
\Big\{
\d\big(\e_{\blq+\blk\n}
-\e_{\blk\m}-\w_{\blq\a}\big)
\Big[\big(f^{\rm el}_{\blq+\blk\n}-1\big)f^{\rm el}_{\blk\m}B^{<}_{\blq\a}-
f^{\rm el}_{\blq+\blk\n}\big(f^{\rm el}_{\blk\m}-1\big)B^{>}_{\blq\a}
\Big]
\nn\\
&+
\d\big(\e_{\blq+\blk\n}
-\e_{\blk\m}+\w_{\blq\a}\big)
\Big[\big(f^{\rm el}_{\blq+\blk\n}-1\big)f^{\rm el}_{\blk\m}B^{>\ast}_{-\blq\a}
-
f^{\rm el}_{\blq+\blk\n}\big(f^{\rm el}_{\blk\m}-1\big)B^{<\ast}_{-\blq\a}
\Big]\Big\}
\label{S21markov}
\end{align}
and 
\begin{align}
S^{22,\rm ph}_{\blq\a\a}&=\p\sum_{\blk\m\n}
\frac{|g_{\blq\a,\m\n}(\blk)|^{2}}{2}
\Big\{
\d\big(\e_{\blq+\blk\n}
-\e_{\blk\m}-\w_{\blq\a}\big)
\Big[\big(f^{\rm el}_{\blq+\blk\n}-1\big)f^{\rm el}_{\blk\m}C^{<}_{\blq\a}
-
f^{\rm el}_{\blq+\blk\n}\big(f^{\rm el}_{\blk\m}-1\big)C^{>}_{\blq\a}
\Big]
\nn\\
&-
\d\big(\e_{\blq+\blk\n}
-\e_{\blk\m}+\w_{\blq\a}\big)
\Big[\big(f^{\rm el}_{\blq+\blk\n}-1\big)f^{\rm el}_{\blk\m}C^{>\ast}_{-\blq\a}
-
f^{\rm el}_{\blq+\blk\n}\big(f^{\rm el}_{\blk\m}-1\big)C^{<\ast}_{-\blq\a}
\Big]\Big\}+{\rm h.c.},
\end{align}
and the more obvious ones
\begin{align}
&S^{11,\rm ph}_{\blq\a\a}=0,
\\
&S^{12,\rm ph}_{\blq\a\a}=[S^{21,\rm phon}_{\blq\a\a}]^{\ast}.
\end{align}

With these preliminary results we can construct the equations of 
motion for the phononic occupations and coherences. From 
Eqs.~(\ref{gammamatele}) we have 
\begin{subequations}
\begin{align}
f^{\rm ph}_{\blq\a}&=\frac{1}{2}\Big[
\w_{\blq\a}\g^{11,<}_{\blq\a\a}+\frac{1}{\w_{\blq\a}}\g^{22,<}_{\blq\a\a}
-i\big(\g^{12,<}_{\blq\a\a}-\g^{21,<}_{\blq\a\a}\big)\Big],
\\
\Th_{\blq\a}&=\frac{1}{2}\Big[
\w_{\blq\a}\g^{11,<}_{\blq\a\a}-\frac{1}{\w_{\blq\a}}\g^{22,<}_{\blq\a\a}
+i\big(\g^{12,<}_{\blq\a\a}+\g^{21,<}_{\blq\a\a}\big)\Big].
\end{align}
\end{subequations}
Using Eq.~(\ref{sepeqgamma}) we then find
\begin{subequations}
\begin{align}
&\frac{d}{dt}f^{\rm ph}_{\blq\a}(t)=S^{\rm ph-occ}_{\blq\a\a}(t),
\\
&\frac{d}{dt}\Th_{\blq\a}(t)+2 i 
\w_{\blq\a}\Th_{\blq\a}=S^{\rm ph-coh}_{\blq\a\a}(t),
\label{eomphcoh}
\end{align}
\end{subequations}
where 
\begin{align}
S^{\rm ph-occ}_{\blq\a\a}&=\frac{1}{2}
\Big[
\w_{\blq\a}S^{11,\rm ph}_{\blq\a\a}+\frac{1}{\w_{\blq\a\a}}S^{22,\rm 
ph}_{\blq\a\a}
-i\big(S^{12,\rm ph}_{\blq\a\a}-S^{21,\rm ph}_{\blq\a\a}\big)\Big]
\nn\\
&=2\p\sum_{\blk\m\n}
\frac{|g_{\blq\a,\m\n}(\blk)|^{2}}{2\w_{\blq\a}}
\Big\{
\d\big(\e_{\blq+\blk\n}
-\e_{\blk\m}-\w_{\blq\a}\big)
\Big[\big(f^{\rm el}_{\blq+\blk\n}-1\big)f^{\rm el}_{\blk\m}f^{\rm ph}_{\blq\a}
-
f^{\rm el}_{\blq+\blk\n}\big(f^{\rm el}_{\blk\m}-1\big)
\big(f^{\rm ph}_{\blq\a}+1\big)\Big]
\nn\\
&+\d\big(\e_{\blq+\blk\n}
-\e_{\blk\m}+\w_{\blq\a}\big)
\Big[\big(f^{\rm el}_{\blq+\blk\n}-1\big)f^{\rm el}_{\blk\m}
-
f^{\rm el}_{\blq+\blk\n}\big(f^{\rm el}_{\blk\m}-1\big)
\Big]\Re\big[\Th_{\blq\a}\big]\Big\},
\label{scattocc}
\end{align}
and
\begin{align}
S^{\rm ph-coh}_{\blq\a\a}&=\frac{1}{2}
\Big[\w_{\blq\a\a}S^{11,\rm ph}_{\blq\a\a}-\frac{1}{\w_{\blq\a\a}}S^{22,\rm 
ph}_{\blq\a\a}
+i\big(S^{12,\rm ph}_{\blq\a\a}+S^{21,\rm ph}_{\blq\a\a}\big)\Big]
\nn\\
&=\p\sum_{\blk\m\n}\frac{|g_{\blq\a,\m\n}(\blk)|^{2}}{2\w_{\blq\a}}
\Big\{
\d\big(\e_{\blq+\blk\n}-\e_{\blk\m}-\w_{\blq\a}\big)
\Big[\big(f^{\rm el}_{\blq+\blk\n}-1\big)
f^{\rm el}_{\blk\m}\big(\Th_{\blq\a}-f^{\rm ph}_{\blq\a}\big)
-f^{\rm el}_{\blq+\blk\n}\big(f^{\rm el}_{\blk\m}-1\big)
\big(\Th_{\blq\a}-f^{\rm ph}_{\blq\a}-1\big)\Big]
\nn \\
&+
\d\big(\e_{\blq+\blk\n}-\e_{\blk\m}+\w_{\blq\a}\big)
\Big[\big(f^{\rm el}_{\blq+\blk\n}-1\big)f^{\rm el}_{\blk\m}
\big(f^{\rm ph}_{-\blq\a}+1-\Th_{\blq\a}\big)
-f^{\rm el}_{\blq+\blk\n}\big(f^{\rm el}_{\blk\m}-1\big)
\big(f^{\rm ph}_{-\blq\a}-\Th_{\blq\a}\big)\Big]\Big\}.
\label{scattcoh}
\end{align}
\end{widetext}

The equation of motion for the phononic coherences deserves further investigation. 
Let us write $\Th_{\blq\a}=\Th^{(r)}_{\blq\a}+i
\Th^{(i)}_{\blq\a}$ as the sum of its real and imaginary part. Then 
Eqs.~(\ref{eomphcoh}) and (\ref{scattcoh}) imply
\begin{subequations}
\begin{align}
&\frac{d}{dt}\Th^{(i)}_{\blq\a}+2 
\w_{\blq\a}\Th^{(r)}_{\blq\a}=-\G^{\rm 
coh}_{\blq\a}\Th^{(i)}_{\blq\a},
\label{eomphimcohi}
\\
&\frac{d}{dt}\Th^{(r)}_{\blq\a}-2 
\w_{\blq\a}\Th^{(i)}_{\blq\a}=-\G^{\rm 
coh}_{\blq\a}\Th^{(r)}_{\blq\a}+\left.
S^{\rm ph-coh}_{\blq\a\a}\right|_{\Th_{\blq\a}=0},
\label{eomphimcohr}
\end{align}
\label{eomphimcoh}
\end{subequations}
where
\begin{align}
\G^{\rm coh}_{\blq\a}&=
\p\sum_{\blk\m\n}\frac{|g_{\blq\a,\m\n}(\blk)|^{2}}{2\w_{\blq\a}}
\nn\\
&\times\left[\d\big(\e_{\blq+\blk\n}-\e_{\blk\m}-\w_{\blq\a}\big)-
\d\big(\e_{\blq+\blk\n}-\e_{\blk\m}+\w_{\blq\a}\big)\right]
\nn\\
&\times\left[f^{\rm el}_{\blk\m}-f^{\rm el}_{\blq+\blk\n}\right].
\end{align}
The {\em coherence rate} $\G^{\rm coh}_{\blq\a}$ is positive 
when the electronic 
occupations satisfy the inequality $f_{\blk\m}>f_{\blq+\blk\n}$ 
for $\e_{\blq+\blk\n}>\e_{\blk\m}$ and 
indices $(\m,\n)$ that are both conduction or 
both valence bands (contributions with $\m$ a valence index and $\n$ a conduction 
index or viceversa vanish due to the Dirac delta); this condition holds true for a 
quasi-thermalized distribution of carriers.
We prove in Section~\ref{steadysubsec} that 
$\left.S^{\rm ph-coh}_{\blq\a\a}\right|_{\Th_{\blq\a}=0}$ vanishes 
for thermal electronic and phononic occupations, and it is therefore 
small for occupations close to thermal ones. Ignoring this term 
in Eq.~(\ref{eomphimcohr}) and assuming $\G^{\rm coh}_{\blq\a}(t)$ weakly 
dependent on time, the most general solution for the phononic 
coherences is
\begin{align}
\Th^{(i)}_{\blq\a}(t)&=\Th_{0,\blq\a}\cos\big(2\w_{\blq\a}t+\f_{0,\blq\a}\big)
e^{-\G^{\rm coh}_{\blq\a}t},
\nn\\
\Th^{(r)}_{\blq\a}(t)&=\Th_{0,\blq\a}\sin\big(2\w_{\blq\a}t+\f_{0,\blq\a}\big)
e^{-\G^{\rm coh}_{\blq\a}t}.
\nn
\end{align}
It is worth remarking that the Markovian approximation of the GKBA 
equations of motion for the phononic coherences differs from  
the MGKBA equation Eqs.~(\ref{eomphimcoh}) in that the sign of 
$\Gamma^{\rm coh}_{\blq\a}$ is reversed, see Appendix~\ref{gkbafailure}.   
This implies that the equilibrium solution is unstable in the GKBA 
version of the SEPE.  

\subsection{Short and long driving}

\begin{center}
\begin{table*}[tbp]
    \centering
    \begin{tabular}{|c|c|}
        \hline
	 & \\
Definition  & Equations   \\
\hline
& \\
Electronic occupation & 
$\frac{d}{dt}f^{\rm el}_{\blk\m}+i\sum_{\n\neq \m}
\Big(\W^{\rm ren}_{\blk\m\n}\,p_{\blk\n\m}-p_{\blk\m\n}\,\W^{\rm 
ren}_{\blk\n\m}\Big)=
S^{\rm el}_{\blk\m\m}$ \\
& \\
\hline
& \\
Electronic polarization &
$\frac{d}{dt}p_{\blk\m\n}+
i\big(\tilde{\e}_{\blk\m}-\tilde{\e}_{\blk\n}\big)p_{\blk\m\n}
+i\W^{\rm ren}_{\blk\m\n}\big(f^{\rm el}_{\blk\n}-f^{\rm el}_{\blk\m}\big)
+i\sum_{\n'\neq \n}
\W^{\rm ren}_{\blk\m\n'}\,p_{\blk\n'\n}-i\sum_{\n'\neq 
\m}p_{\blk\m\n'}\,\W^{\rm ren}_{\blk\n'\n}=
-\G^{\rm pol}_{\blk\m\n} p_{\blk\m\n}$
\\
& \\
\hline
& \\
Nuclear displacement & $\frac{d^{2}}{dt^{2}}U_{{\mathbf 0}\a}
+\w^{2}_{{\mathbf 0}\a}U_{{\mathbf 0}\a}
=-\sum_{\blk\m\n}g_{{\mathbf 
0}\a,\n\m}(\blk)\big[\d_{\m\n}\D f^{\rm el}_{\blk\m}+
p^{\rm intra}_{\blk\m\n}\big]
-\int dt'\sum_{\subalign{&\blk\blk'\\&\m\m'\\ 
&\n\n'}}g_{{\mathbf 
0}\a,\n\m}(\blk)\chi^{\rm 
R}_{\blk\blk',\subalign{&\m\m'\\\ 
&\n\n'}}({\mathbf 
0};t,t')\W_{\blk'\m'\n'}(t')$ \\
 & \\
\hline
& \\
Phonon occupation & 
$\frac{d}{dt}f^{\rm ph}_{\blq\a}=S^{\rm ph-occ}_{\blq\a\a}$\\
 & \\
\hline
& \\
Phonon coherence & 
$\frac{d}{dt}\Th_{\blq\a}+2\iu 
\w_{\blq\a}\Th_{\blq\a}=S^{\rm ph-coh}_{\blq\a\a}$\\
 & \\
\hline
& \\
Quasiparticle energy &   
$\tilde{\e}_{\blk\m}(t)=\e_{\blk\m}+\sum_{\blk'\m'}
(v_{\blk\m\blk'\m'\blk'\m'\blk\m}-W_{\blk\m\blk'\m'\blk\m\blk'\m'})
[f_{\blk'\m'}(t)-f_{\blk'\m'}(0)]$
\\
 & \\
\hline
& \\
Renorm. Rabi frequency & $\W^{\rm ren}_{\blk\m\n}=\W_{\blk\m\n}+
\sum_{\a}g_{{\mathbf 0}\a,\m\n}(\blk)
U_{{\mathbf 0}\a}$
\\
& \\
\hline
\end{tabular}
    \caption{Table summarizing the SEPE.}
    \label{sepeqtab}
\end{table*}
\end{center}

In Table~\ref{sepeqtab} we summarize the SEPE for the 
electronic and phononic degrees of freedom. 
The electronic 
scattering term is given by the sum of Eq.~(\ref{sbescattdiagmarkov}) 
-- $GW$ -- and 
Eq.~(\ref{sbescattdiagFMgenmarkov}) -- Fan-Migdal -- 
whereas the phononic scattering terms are 
given by Eqs.~(\ref{scattocc}) and (\ref{scattcoh}). All scattering 
terms 
except the $GW$ one depend on both phononic occupations and coherences. 

If the duration $T_{\rm drive}$
of the  driving field is 
much shorter than a typical phonon period (for a phonon frequency  
$\lesssim 100$~meV the phonon period is $\gtrsim 40$~fs) 
then the nuclei remain essentially still while 
the field is on. For such short drivings we can neglect the last term 
in the third equation of Table~\ref{sepeqtab},
as it grows linearly with $T_{\rm drive}$. The 
energy shift $\sum_{\a}g_{{\mathbf 0}\a,\m\n}(\blk)U_{{\mathbf 0}\a}$ 
in $\W^{\rm ren}_{\blk\m\n}$, see Eq.~(\ref{renrabi}), is typically 
of the order of a 
few meV and it is responsible for time-dependent modulations in 
the optical spectra~\cite{trovatello_strongly_2020,li_single-layer_2021,mor_photoinduced_2021,jeong_coherent_2016,sayers_strong_2023}.
To capture this effect it is crucial to use  $\W^{\rm ren}_{\blk\m\n}$ 
in the two terms that multiply the polarizations
in the second equation of Table~\ref{sepeqtab}. 
All other $\W^{\rm ren}_{\blk\m\n}$  
can be approximated with $\W_{\blk\m\n}$.  

For long driving fields, i.e.,  $T_{\rm drive}$ a few hundreds of fs or 
longer, the last term in the third equation of 
Table~\ref{sepeqtab} cannot be discarded. In this case the 
nuclei are expected to slowly attain new positions and no 
time-dependent modulations of the optical spectra are to be expected. 
The nuclear shifts are mainly responsible for 
a few meV renormalization of the quasi-particle energies.  
Thus, if our focus is solely on occupations and coherences, or if we 
do not require a meV resolution of the optical spectra then  
we can solve the SEPE with $U_{{\mathbf 0}\a}=0$.

\subsection{Equilibrium and steady-state solutions}
\label{steadysubsec}

Let us discuss the stationary solutions of the SEPE. 
We assign a chemical potential $\m_{c}$ to all 
conduction bands and $\m_{v}$ to all valence bands, respectively, 
and show that all scattering terms vanish if the electronic occupations 
$f^{\rm el}_{\blk\n}=1/[e^{\b(\e_{\blk\n}-\m_{\n})}+1]$ 
(noninteracting finite-temperature  
electrons), the phononic occupations 
$f^{\rm ph}_{\blq\a}=1/[e^{\b\w_{\blq\a}}-1]$
(noninteracting 
finite-temperature phonons), and the phononic coherences 
$\Th_{\blq\a}=0$.
Let us consider the $GW$ scattering 
term in Eq.~(\ref{sbescattdiagmarkov}). Taking into account that 
$f^{\rm el}_{\blk\n}/[1-f^{\rm el}_{\blk\n}]=e^{-\b(\e_{\blk\n}-\m_{\n})}$,
the energy conservation enforced by the Dirac delta implies
\begin{align}
\frac{f^{\rm el}_{\blk+\blp\n}f^{\rm el}_{\blq\m'}e^{\b(\m_{\m'}+\m_{\n})}}
{(f^{\rm el}_{\blk+\blp\n}-1)(f^{\rm el}_{\blq\m'}-1)}
=\frac{f^{\rm el}_{\blq+\blp\n'}f^{\rm el}_{\blk\m}e^{\b(\m_{\n'}+\m_{\m})}}{
(f^{\rm el}_{\blq+\blp\n'}-1)(f^{\rm el}_{\blk\m}(t)-1)},
\nn
\end{align}
and therefore all terms in $S_{\blk\m\m}^{\rm GW}$ with 
$\m_{\m'}+\m_{\n}=\m_{\n'}+\m_{\m}$ vanish. The only case for which 
$\m_{\m'}+\m_{\n}\neq \m_{\n'}+\m_{\m}$ is when $(\m',\n)$ are both 
conduction (valence) bands and $(\n',\m)$ are both valence 
(conduction) bands. However, for this choice of indices 
the argument of the Dirac delta is at least twice the quasi-particle gap. 
We conclude that 
$S_{\blk\m\m}^{\rm GW}=0$.

Similarly, for the 
Fan-Migdal scattering term in Eq.~(\ref{sbescattdiagFMgenmarkov}) we have
\begin{align}
\frac{f^{\rm el}_{\blk\m}}{f^{\rm el}_{\blk\m}-1}
[f^{\rm ph}_{\blk-\blq\a}+1]
=e^{-\b(\m_{\n}-\m_{\m})}\frac{f^{\rm el}_{\blq\n}}{f^{\rm el}_{\blq\n}-1}
f^{\rm ph}_{\blk-\blq\a},
\nn
\end{align}
where we enforce the energy conservation 
$\e_{\blk\m}=\e_{\blq\n}+\w_{\blk-\blq\a}$. 
A similar relation can be derived for the term with 
$\e_{\blk\m}=\e_{\blq\n}-\w_{\blk-\blq\a}$.
Therefore all terms in $S_{\blk\m\m}^{\rm FM}$ with 
$\m_{\n}=\m_{\m}$ vanish. The only case for which 
$\m_{\n}\neq \m_{\m}$ is when $\n$ is a
conduction (valence) band and $\m$ is a valence 
(conduction) band. However, for this choice of indices 
the argument of the Dirac delta is about the quasi-particle gap. 
We conclude that 
$S_{\blk\m\m}^{\rm FM}=0$. The same arguments can be used to show 
that $S^{\rm ph-occ}_{\blq\a\a}=S^{\rm ph-coh}_{\blq\a\a}=0$. 

The most general steady-state 
solution of the SEPE
with $\W^{\rm ren}_{\blk\m\n}=0$ is given by noninteracting electronic and 
phononic occupations {\em at the same temperature}, and vanishing 
electronic polarizations and phononic coherences. Among all these 
solutions there exists the equilibrium one, where $\b$ is the 
equilibrium inverse 
temperature and $\m_{c}=\m_{v}=\m$ is the equilibrium chemical 
potential. The steady-state solution, if attained, is expected to 
have a temperature higher than the 
equilibrium temperature since the external field injects 
energy in the system. This means that $\D 
f^{\rm el}_{\blk\m}=f^{\rm el}_{\blk\m}(t\to\iif)-f^{\rm el}_{\blk\m}(t=0)$ is, in general, 
different from zero, and therefore the nuclei attain new positions
$U_{{\mathbf 0}\a}(t\to\iif)=-\frac{1}{\w_{{\mathbf 0}\a}^{2}}
\sum_{\blk\m}g_{{\mathbf 
0}\a,\m\m}(\blk)\D f^{\rm el}_{\blk\m}$, see Eq.~(\ref{sducp2}).
These displacements can be either negative or positive, depending on 
the sign and magnitude of the $e$-$ph$ couplings.
Because of the non-zero nuclear displacements, $\W^{\rm 
ren}_{\blk\m\n}(t\to\iif)$ is small but not exactly zero. 
Consequently, the steady-state occupations, polarizations, and 
coherences differ slightly from the thermal values.

\section{Recovering the Semiconductor Bloch and Boltzmann equations}
\label{recoverysec}

Historically, the SBE exclusively addressed 
the electron dynamics~\cite{haug_quantum_1994,schubert_sub-cycle_2014}, 
implicitly assuming that the phonons remained in 
thermal 
equilibrium~\cite{marini_competition_2013,steinhoff_nonequilibrium_2016,MolinaSanchez2017}. These equations follow from the SEPE by setting 
$U_{{\mathbf 0},\a}=\Th_{\blq\a}=0$ and $f^{\rm 
ph}_{\blq\a}=1/[e^{\b\w_{\blq\a}}-1]$, thus only the first two 
equations in Table~\ref{sepeqtab} need to be solved. Improvements of 
the SBE in which the phononic occupations $f^{\rm 
ph}_{\blq\a}$ satisfy their own equation of 
motion (fourth equation in Table~\ref{sepeqtab}) have also been 
considered~\cite{malic_microscopic_2011}.  

The SBE simplify further in the so 
called {\em incoherent regime}. For times long after the 
perturbation caused by the external field it is reasonable to assume
that the system is well 
described by a many-body density matrix of the form
$\hat{\r}(t)=
\sum_{k}w_{k}(t)|\Q_{k}\ket\bra\Q_{k}|$,
where the many-body states 
$|\Q_{k}\ket$ have a well defined number of electrons and phonons or, 
equivalently, are eigenstates of the electronic and phononic number 
operators
\begin{subequations}
\begin{align}
\hat{n}^{\rm el}_{\blk\m}|\Q_{k}\ket&=\hat{d}^{\dag}_{\blk\m}\hat{d}_{\blk\m}|\Q_{k}\ket
=n^{\rm el}_{\blk\m}|\Q_{k}\ket,\quad\quad n^{\rm el}_{\blk\m}=0,1
\\
\hat{n}^{\rm 
ph}_{\blq\a}|\Q_{k}\ket&=\hat{b}^{\dag}_{\blq\a}\hat{b}_{\blq\a}|\Q_{k}\ket
=n^{\rm ph}_{\blq\a}|\Q_{k}\ket,\quad\quad n^{\rm 
ph}_{\blq\a}=0,1,2,\ldots
\end{align}
\label{numbereigenhyp}
\end{subequations}
In the incoherent regime we have $p_{\blk\m\n}(t)=U_{{\mathbf 
0}\a}=\Th_{\blq\n}=0$, thus only the equations of motion for 
$f^{\rm el}_{\blk\m}$ and $f^{\rm ph}_{\blq\a}$ (first and fourth 
equations in Table~\ref{sepeqtab}) need to be solved. These are the 
BE~\cite{jauho-book,sadasivam_theory_2017,ono_ultrafast_2020,tong_toward_2021,caruso_nonequilibrium_2021,caruso_ultrafast_2022}.
The BE do not account for the interaction between the
electrons and the driving field. The  
nonequilibrium distribution of electrons and phonons enters as the 
initial value of $f^{\rm el}_{\blk\m}$ and $f^{\rm ph}_{\blq\a}$.

We conclude by observing that the SBE 
and BE can be derived from the GKBA
only if we {\em impose} that $\Th_{\blq\a}=0$. 
In fact, $\Th_{\blq\a}=0$ is not a stable solution in GKBA, see 
Appendix~\ref{gkbafailure}.


\section{Conclusions}
\label{conclsec}

Based on our recent work on the {\em ab initio} many-body theory of 
electrons and phonons~\cite{stefanucci_in-and-out_2023} 
we derive a simplified set of equations of 
motion for the electronic occupations and polarizations, nuclear 
displacements as well as phononic occupations and coherences, 
highlighting all the underlying simplifications.
By explicitly including the laser field, it is possible to 
create a nonequilibrium population of electrons and phonons, 
while simultaneously transferring the laser coherence to both electrons and the 
nuclear lattice.  
Currently available electronic structure codes can be used to 
calculate the screened Coulomb interaction $W$ and $e$-$ph$ coupling 
$g$, which are the only necessary ingredients to  run first-principles 
SEPE simulations. In particular, 
the SEPE equations can be readily implemented in SBE and BE codes 
through a minor change in the Fan-Migdal scattering term, i.e., 
$f^{\rm ph}_{\blq}\to f^{\rm ph}_{\blq}+\Th_{\blq\a}$, and by 
adding the equations of motion for the nuclear displacements and 
phononic coherences. These new features pave the way for 
first-principles studies of the coupling between coherent phonons and 
excitons as well as squeezed phonon states and time-dependent 
Debye-Waller factors. Whether the consistent 
treatment of phononic occupations and coherences has also an impact 
on the coherent-to-incoherent crossover~\cite{Lloyd_Hughes_2021,perfetto_real-time_2023} and on the thermalization of 
electrons and phonons remain to be seen case by case.

On a more fundamental level our work sheds light on the SBE and BE as 
it clarifies how to derive these methods from the {\em ab initio} KBE. 
The mirrored version of the GKBA has been essential in this endeavour. In fact, the GKBA 
yields unphysical polarization rates and an exponentially diverging 
solution for the phononic coherences. On the contrary, in MGKBA the equations of 
motion for the electronic and phononic density matrices have  
the same structure as a rate equation, and the rates naturally turn 
out to be positive once the Markovian approximation is made.

We conclude by outlining two important future directions. Following 
the strategy presented in this work it would be interesting to derive the scattering 
terms arising from anharmonic effects, e.g., the $ph$-$ph$ 
interaction~\cite{ziman_electrons_1960}, which are expected play a 
role in the thermalization of the 
lattice~\cite{caruso_nonequilibrium_2021}. A second 
important direction is the derivation of an effective equation for 
the exciton-phonon 
dynamics~\cite{chen_exciton-phonon_2020,cudazzo_first-principles_2020,antonius_theory_2022,paleari_exciton-phonon_2022}.  
We anticipate the appearance of phononic coherences 
in both cases.

\begin{acknowledgments}
G.S. and E.P. acknowledge the financial support from MIUR PRIN (Grant No. 
20173B72NB), MIUR PRIN (Grant No. 2022WZ8LME), INFN
through the TIME2QUEST project and Tor Vergata University through 
Project TESLA.
\end{acknowledgments}

\appendix

\section{Response function in the Ehrenfest approximation}
\label{ehchisec}

Let us denote by $\tilde{\chi}$ the full response function of the 
electron-phonon system. The diagrammatic expansion of $\tilde{\chi}$ 
contains diagrams with both $e$-$e$  and $e$-$ph$ 
interactions. To first order in the external field $\W$ the change $\D\r$ 
in the electronic density matrix is given by (omitting the dependence 
on momenta)
\begin{align}
\D\r_{\m\n}(t)=\!\int \!dt'\!\sum_{\m'\n'}\tilde{\chi}^{\rm 
R}_{\subalign{&\m\m'\\\ 
&\n\n'}}(t,t')\W_{\m'\n'}(t').
\label{deltarho}
\end{align}
The Ehrenfest approximation to the response function is 
\begin{align}
\tilde{\chi}^{\rm 
R}_{\subalign{&\m\m'\\\ 
&\n\n'}}(t,t')=\chi^{\rm 
R}_{\subalign{&\m\m'\\\ 
&\n\n'}}(t,t')
+\sum_{\subalign{&\a\b\\
&\m\m'\\&\n\n'}}
\int dt_{1}dt_{2}
\chi^{\rm 
R}_{\subalign{&\m\r\\\ 
&\n\s}}(t,t_{1})g_{\a,\r\s}
\nn\\
\times D^{11,\rm R}_{\a\b}(t_{1},t_{2})
g_{\b,\s'\r'}\tilde{\chi}^{\rm 
R}_{\subalign{&\r'\m'\\\ 
&\s'\n'}}(t_{2},t'),
\label{ehappchi}
\end{align}
where $\chi$ is the response function at clamped nuclei.
Substituting Eq.~(\ref{ehappchi}) into Eq.~(\ref{deltarho}) and 
taking into account that~\cite{stefanucci_in-and-out_2023} 
\begin{align}
U_{\a}(t_{1})=\sum_{\b\r'\s'}\int dt_{2}	D^{11,\rm R}_{\a\b}(t_{1},t_{2})
g_{\a,\s'\r'}\D\r_{\r'\s'}(t_{2}),
\end{align}
we find Eq.~(\ref{lrfehdyn}).

\section{Markovian limit of the GKBA equations}
\label{gkbafailure}

In GKBA the electronic and phononic lesser GFs  
become [compare with Eqs.~(\ref{gkbaapprox}), (\ref{D11<gkba}) and (\ref{D12<gkba})]
\begin{align}
G^{<}_{\blk\m\m}(t,t')=i
e^{-i \e_{\blk\m}(t-t')}
\big[
\th(t-t')f^{\rm el}_{\blk\m}(t')+\th(t'-t)f^{\rm el}_{\blk\m}(t)\big].
\label{gkbaapprox1}
\end{align}
\begin{align}
D^{11,\lessgtr}_{\blq\a\a}(t,t')&=\frac{\th(t-t')}{2i\w_{\blq\a}}
\big[B^{\lessgtr}_{\blq\a}(t')e^{-\iu\w_{\blq\a}(t-t')}+
B^{\gtrless\ast}_{-\blq\a}(t')
e^{\iu\w_{\blq\a}(t-t')}\big]
\nn\\
+&\frac{\th(t'-t)}{2i\w_{\blq\a}}
\big[B^{\lessgtr\ast}_{\blq\a}(t)e^{-\iu\w_{\blq\a}(t-t')}+
B^{\gtrless}_{-\blq\a}(t)e^{\iu\w_{\blq\a}(t-t')}\big],
\label{D11<gkba1}
\end{align}
\begin{align}
D^{12,\lessgtr}_{\blq\a\a}(t,t')&=\frac{\th(t-t')}{2}
\big[C^{\lessgtr}_{\blq\a}(t')e^{-\iu\w_{\blq\a}(t-t')}-
 C^{\gtrless\ast}_{-\blq\a}(t')e^{\iu\w_{\blq\a}(t-t')}\big]
\nn\\
+&\frac{\th(t'-t)}{2}
\big[
B^{\lessgtr\ast}_{\blq\a}(t)e^{-\iu\w_{\blq\a}(t-t')}-
B^{\gtrless}_{-\blq\a}(t)e^{\iu\w_{\blq\a}(t-t')}\big],
\label{D12<gkba1}
\end{align}
The expression for $G^{>}_{\blk\m\n}(t,t')$ is identical provided that 
$f^{\rm el}_{\blk\m}\to 
f^{\rm el}_{\blk\m}-1$, see (\ref{commreln<>}).
Implementing the same simplifications leading to the SEPE we obtain 
the same equations as in Table~\ref{sepeqtab} but with different 
phononic scattering  terms (the electronic scattering terms $S^{\rm 
GW}$ and $S^{\rm FM}$ remain unchanged). 
In particular the scattering terms for the phononic occupations and 
coherences read [compare with Eqs.~(\ref{scattocc}) and (\ref{scattcoh})]
\begin{widetext}
\begin{align}
S^{\rm ph-occ}_{\blq\a}&=
2\p\sum_{\blk\m\n}
\frac{|g_{\blq\a,\m\n}(\blk)|^{2}}{2\w_{\blq\a}}
\Big\{
\d\big(\e_{\blq+\blk\n}
-\e_{\blk\m}-\w_{\blq\a}\big)
\nn\\
&\times
\Big[\big(f_{\blq+\blk\n}-1\big)f_{\blk\m}
\big(f^{\rm ph}_{\blq\a}+\Re\big[\Th_{\blq\a}\big]\big)
-
f_{\blq+\blk\n}\big(f_{\blk\m}-1\big)\big(f^{\rm 
ph}_{\blq\a}+1+\Re\big[\Th_{\blq\a}\big]\big)\Big\},
\end{align}
\begin{align}
S^{\rm ph-coh}_{\blq\a}&=
2\p\sum_{\blk\m\n}\frac{|g_{\blq\a,\m\n}(\blk)|^{2}}{2\w_{\blq\a}}
\Big\{-
\d\big(\e_{\blq+\blk\n}-\e_{\blk\m}-\w_{\blq\a}\big)
\Big[\big(f_{\blq+\blk\n}-1\big)f_{\blk\m}\big(f^{\rm ph}_{\blq\a}+\Th_{\blq\a}\big)
-f_{\blq+\blk\n}\big(f_{\blk\m}-1\big)\big(f^{\rm 
ph}_{\blq\a}+1+\Th_{\blq\a}\big)\Big]
\nn\\
&+
\d\big(\e_{\blq+\blk\n}-\e_{\blk\m}+\w_{\blq\a}\big)
\Big[\big(f_{\blq+\blk\n}-1\big)f_{\blk\m}\big(f^{\rm 
ph}_{-\blq\a}+1+\Th_{\blq\a}\big)
-f_{\blq+\blk\n}\big(f_{\blk\m}-1\big)\big(f^{\rm 
ph}_{-\blq\a}+\Th_{\blq\a}\big)\Big]\Big\}.
\end{align}
\end{widetext}
The equation of motion for the real and imaginary part of the 
coherences are identical to Eqs.~(\ref{eomphimcoh}), but the sign of 
$\G^{\rm coh}_{\blq\a}$ is reversed.

The GKBA poses issues when employed to estimate polarization rates as well.
In MGKBA the Fan-Migdal contribution to 
$-\G^{\rm el,>}_{\blk\m\m}p_{\blk\m\n}$, see Eqs.~(\ref{sbescattpol}) and 
(\ref{polscattrate}), is calculated with the polarization rate
\begin{align}
\G^{\rm el,>}_{\blk\m\m}=2\p
\sum_{\blq\n'\a}\frac{\big|g_{\blq-\blk\a,\m\n'}(\blk)\big|^{2}}{\w_{\blk-\blq\a}}
\d\big(
\e_{\blq\n'}-\e_{\blk\m}
+\w_{\blk-\blq\a}
\big)
\nn\\
\times\big(1-f_{\blq\n'}\big)\Re\big[B^{>}_{\blk-\blq\a}\big].
\end{align}
For the argument  of the Dirac delta to vanish the index $\n'$ must 
belong to the same ``class'' (conduction or valence) as the index 
$\m$ and therefore the polarization rate is dominated by either 
conduction-conduction or valence-valence $e$-$ph$ couplings.
In GKBA the same 
term is replaced by $-\G^{\rm el,>}_{\blk\m\n}p_{\blk\m\n}$ with 
\begin{align}
\G^{\rm el,>}_{\blk\m\n}=2\p
\sum_{\blq\n'\a}\frac{\big|g_{\blq-\blk\a,\m\n'}(\blk)\big|^{2}}{\w_{\blk-\blq\a}}
\d\big(
\e_{\blq\n'}-\e_{\blk\n}
+\w_{\blk-\blq\a}
\big)
\nn\\
\times\big(1-f_{\blq\n'}\big)\Re\big[B^{>}_{\blk-\blq\a}\big],
\end{align}
i.e., the argument of the Dirac delta is calculated with 
$\e_{\blk\n}$ instead of $\e_{\blk\m}$. Using the same reasoning,  we 
infer that the polarization rate is dominated by either 
conduction-valence or valence-conduction $e$-$ph$ couplings, which is 
not to be expected. Moreover, there is no guarantee that the matrix 
$\G^{\rm el,>}_{\blk\m\n}$ is positive semi-definite for quasi-thermal 
distributions.


%

\end{document}